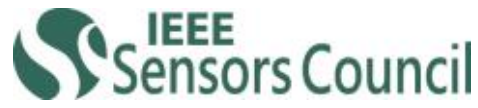

# Adaptive Channel Hopping for IEEE 802.15.4 TSCH-Based Networks: A Dynamic Bernoulli Bandit Approach

Nastooh Taheri Javan, *Senior Member*, *IEEE*, Masoud Sabaei, and Vesal Hakami

***Abstract*—In IEEE 802.15.4 standard for low-power low-range wireless communications, only one channel is employed for transmission which can result in increased energy consumption, high network delay and poor packet delivery ratio (PDR). In the subsequent IEEE 802.15.4-2015 standard, a Time-slotted Channel Hopping (TSCH) mechanism has been developed which allows for a periodic yet fixed frequency hopping pattern over 16 different channels. Unfortunately, however, most of these channels are susceptible to high-power coexisting Wi-Fi signal interference and to possibly some other ISM-band transmissions. This interference manifests itself in the form of the presence/absence of other devices with either or both static and dynamic channel selection policies. In order to isolate channels with undesirable conditions, blacklisting mechanisms are defined to adapt the channel hopping process. However, the existing solutions which form blacklists unrealistically assume that the statistical model of the external interference remains fixed, and do not vary over time. In this paper, we realistically assume that the impact of external interferes on 802.15.4 may generally follow a non-stationary pattern, and accordingly formulate the adaptive channel hopping problem as a Dynamic Multi-Armed Bernoulli Bandit (Dynamic MABB) process from the machine learning theory. We then propose an online learning algorithm with track-ability properties for computing an adaptive hopping policy. Simulations confirm that when the statistics of the external interference has a switching regime, the proposed solution outperforms the previous schemes in terms of both energy efficiency as well as two important KPIs for TSCH-based networks, i.e., PDR and latency.**

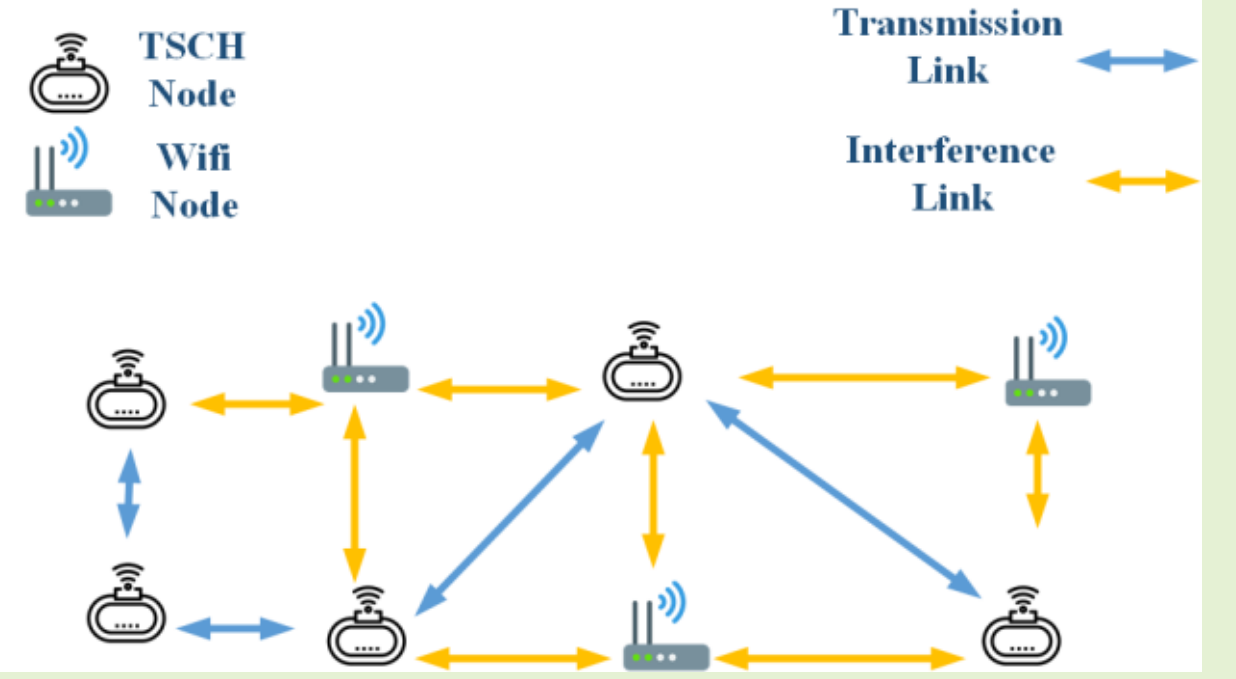

***Index Terms*— IEEE 802.15.4, TSCH, Channel Hopping, Blacklisting, Dynamic MABB Problem.**

## I. Introduction

In recent years, the concept of the Internet of Things (IoT) has been evolving rapidly, making it increasingly possible to connect any device to the Internet (as well as to each other) [1]. One basic requirement in the IoT connectivity landscape is to ensure that nodes are reliably connected to the Internet [2]. Given this key demand, IoT-based settings are expected to satisfy the required specifications mandated by the standards such as IEEE 802.15.4 [3] (which imposes a set of protocols for the provision of two-way multi-hop interconnections between devices with limited power), and LoRa [4] (which mainly defines Medium Access Control (MAC) layer and message formats for single-hop device-to-BS long distance communications).

IEEE 802.15.4 is a popular standard that specifies the physical and MAC layer for low-power low data rate communications [5]. However, it is a single-channel standard in which communications can be heavily affected by interference and fading, and thus incur high delay and a low packet delivery ratio [6]. IEEE 802.15.4-2015 [12] was later introduced as an improved version to IEEE 802.15.4e, presenting a new *time-slotted channel hopping* (TSCH) operating mode to ensure higher reliability by using several channels [7]. In TSCH, at the time of sending a packet over a channel, if the desired channel is affected by destructive factors such as interference, it is possible to resend it on a different channel at another time-slot (possibly in another slot-frame cycle). More specifically, TSCH employs a deterministic periodic hopping pattern on 16 different channels [8]. A drawback with default TSCH operation is that most of its channels are susceptible to interfering transmissions from the 2.4GHz ISM band technologies such as IEEE 802.11 (Wi-Fi) and Bluetooth [9].

Also, in TSCH networks not only can the interference from the uncoordinated neighboring Wi-Fi (or Bluetooth) networks be problematic, multipath fading can also pose problems, e.g. reduced network reliability [10]. It should be noted however that the occupancy of the different frequency channels by the surrounding radio signals is not evenly balanced. Some ISM sub-bands are less occupied or jammed than others, and this diversity should be adaptively exploited to intelligently hop over channels and make a better use of the spectrum.

(Corresponding author: Vesal Hakami.)

N. T. Javan and M. Sabaei are with the Computer Engineering Department, Amirkabir University of Technology (Tehran Polytechnic), Tehran 15875-4413, Iran (e-mail: nastooh@aut.ac.ir; sabaei@aut.ac.ir).

V. Hakami is with the Center of Excellence in Future Networks, School of Computer Engineering, Iran University of Science and Technology, Tehran 16846-13114, Iran (e-mail: vhakami@iust.ac.ir).

Obviously, to provide for such adaptability, one needs to go beyond default TSCH (which offers a fixed hopping pattern).

Adaptive channel hopping in TSCH-based networks can be realized through blacklisting mechanisms [11]. Blacklisting is a technique to omit channels with undesirable conditions from the hopping list. As the statistics of absence/presence of interference sources are neither usually pre-determined nor are their behavior constant, a spatiotemporal prediction is not possible, and instead, blacklisting policies should be computed on-the-fly using model-free learning procedures.

Recently, several blacklisting schemes have been introduced for enabling adaptive TSCH channel hopping. We present an exhaustive review of the related work in Section II.C. However, one common assumption in the existing blacklisting schemes is that the probabilistic model of external interference is stationary in the sense that the statistical properties of the external interference is time-invariant [15-17]. This assumption rarely holds in practice as real-life communications often involve underlying processes that are dynamically evolving [18]. In fact, the presence/absence of external users associated with other wireless technologies do not necessarily exhibit a nice stationary statistical property; also, each non-802.15.4 protocol indeed uses its own rules for channel selection, user partitioning, etc., and all this leads to a hectic use of the spectrum that cannot be expressed in terms of a time-invariant probabilistic model, and has to be learnt and tracked dynamically.

In a departure from the previous works, in this paper, we explore a more realistic approach and generalize the idea of adaptive channel hopping to non-stationary settings. Prior work has mainly utilized a so-called stochastic "multi-armed bandit" (MAB) model [19] to frame the adaptive channel hopping problem (e.g., [20], [21]). MAB is a formalism from the machine learning theory and can be used as a systematic way for a decision-maker to choose among multiple options with uncertain rewards. In particular, the classical MAB describes a setting in which a gambler has to operate a slot machine with multiple arms (levers). The gambler has to decide which arm to play, how many times to play each arm and in which order to play them, and whether to continue with the current arm or try a different arm. In MAB, each arm provides a random reward from an unknown probability distribution specific to that arm. The objective of the gambler is to maximize the sum of rewards earned through a sequence of arm pulls. In modeling the adaptive channel hopping problem as a MAB, the available physical channels can be considered as arms and average PDR as the optimization objective.

In this paper, we formulate the adaptive channel hoping problem in TSCH-networks as a *Dynamic Multi-Armed Bernoulli Bandit* (Dynamic MABB) process [22] which is one of the many non-stationary variants [23] of the classical MAB specifically tailored for MABs with Bernoulli rewards/penalties. In the context of our problem, the sequence of rewards/penalties gained from each arm (channel) indeed forms a Bernoulli process (corresponding to successful/unsuccessful packet transmissions) with an unknown distribution. Also, given the time-varying nature of the external interference, the underlying Bernoulli process would be non-stationary, thereby making Dynamic MABB a very much expressive framework for our case.

In MAB, an online learning algorithm needs to be deployed to tackle with a crucial tradeoff at each trial which is between *exploitation* of the arm that has the highest estimated payoff and *exploration* to get more information about the payoffs of the other arms. In stochastic MAB, some popular learning algorithms include $\varepsilon$-Greedy [24], Upper Confidence Bound (UCB) [25], and Thompson Sampling (TS) [26]. However, such algorithms cannot be justifiably employed for Dynamic MABB as they use sample means to estimate the expected reward of each arm. In stationary MABs, given that the sequence of rewards is i.i.d., forming sample means is sensible from a theoretical perspective, and one could invoke various asymptotic results (e.g., law of large numbers, central limit theorem, etc.) as justification. However, when the underlying distribution changes significantly with time, sample mean-based estimation is not theoretically valid, and can also result in performance bottleneck from a practical viewpoint [27].

Armed with this understanding, in this paper, we consider deploying a different learning algorithm (the so-called AFF-OTS from [22]) which uses an estimator with track-ability properties and is inspired from the adaptive filtering theory [27]. The key idea behind adaptive estimation is to track a time evolving data stream by gradually reducing the weight on older data as new data arrives. There are many learning procedures proposed for arm selection in non-stationary MAB settings (e.g., [28, 29 and 30]), but the recent AFF-OTS procedure proposed in [22] is particularly suited for implementation in an IoT setting as it is a lightweight algorithm, requires very little tuning effort, is quite robust to tuning parameters and unlike prior work in Dynamic MABB, its initialization does not require knowledge about the model structure in advance.

Another notable point is that channel hopping in a TSCH-based network takes influence from the underlying scheduling mechanism. Nonetheless, the standard only provides a framework, without actually mandating a specific scheduling mechanism for time and frequency slot allocation [31]. As such, several interesting scheduling algorithms have been proposed to fill the void (e.g., [31], [54], [55]). In the default single-offset scheduling, however, only one channel is offered to each link in every time-slot [32]. This is while an adaptive channel hopping scheme needs to be built on a multi-offset schedule in which several offsets can be offered to each node at each time-slot. Realizing this and using graph-theoretical algorithms, we use a multi-offset version of our scheduling scheme in [31] that allows each node to select the most suitable offset based on the proposed learning procedure.

In sum, our contributions are as follows:

- We formulate the problem of adaptive channel hopping on top of a multi-offset TSCH schedule using the dynamic (non-stationary) MAB formalism [33] with the goal of maximizing the average PDR. Our formulation is more realistic compared to prior work in that it accounts for the possible non-stationarity of the external interference. Based on recent results in MAB theory [22], we deploy a learning algorithm for computing the channel selection policy. Our algorithm is *suitable for TSCH-based networks susceptible to non-stationary external interference* as it has the capability of tracking the time-varying statistical model of the interference. Also, since our algorithm works based on a number of low cost recursive update formula, it is particularly lightweight with few parameters to adjust, making it suitable for IoT settings.

- Through simulations, we experimentally evaluate our proposed scheme against default TSCH [7] as well as some prior work which assumes stationary interference [21]. Also, a small physical setup is implemented to showcase the memory footprint of the proposed algorithm.

The rest of the paper is organized as follows: in Section II, background and motivation of the paper are described and some relevant studies on adaptive frequency channel hopping for TSCH networks are reviewed. In Section III, our system model including the assumptions is described. In Section IV, the problem is formulated on the basis of Dynamic MABB framework and in Section V the proposed adaptive channel hopping mechanism is presented. In Section VI, our simulation results are presented and the proposed scheme is evaluated against some baseline schemes. In Section VII, we explain a physical setup for evaluating the memory footprint of the proposed algorithm. Finally, Section VIII concludes the paper with highlights of its main results.

## II. Background

In this section, we first give a brief overview of default TSCH channel hopping, and then discuss the rationale behind adaptive hopping. We also discuss a number of related studies, highlight the research gap and motivate our idea in this paper. For ease of reference, important acronyms are summarized in Table I.

### A. Default channel hopping in TSCH

In order to combat internal interference in TSCH-based networks, the nodes use a simple, blind and periodic pattern for frequency hopping in which all the channels are uniformly selected. At each time-slot, a node maps the channel offset into a physical frequency according to the following equation:

$$Frequency = MAP\,[(ASN + Ch_Off) \bmod nCh] \tag{1}$$

where $ASN$ is the network's absolute sequence number of the time-slot, $nCh$ is the number of available channels, $Ch_Off$ is the current scheduled channel offset, and $MAP$ is a bijective function mapping an integer between 0 and $nCh$-1 into the frequency channel. The standard provides a two-dimensional scheduling scheme in which each cell corresponds to a pair of time-slot and channel offset. A TSCH schedule allocates a set of cells to each radio link. It is clear that by employing this blind channel hopping, all the channels are uniformly picked up irrespective of their quality.

### B. Adaptive Channel Hopping

In real environments, signal attenuations (e.g., due to multipath fading and co-existing transmissions) increase the rate of packet transmission failure, thereby leading to a high number of retransmission attempts and energy wastage [9]. In contrast to the neutral round-robin hopping pattern of the default TSCH, an adaptive hopping algorithm constantly tries to detect and avoid undesirable channels with low packet delivery ratio, effectively increasing the energy efficiency.

Adaptive frequency channel hopping schemes can be classified as being either *model-based* or *model-free*. Model-based schemes are built on the assumption that a statistical model of the channel dynamics is known at design time (i.e., prior to actual network deployment and operation). The model-free schemes, on the other hand, are typically based on machine learning algorithms in which a network node starts from a basically zero knowledge of the network dynamics, and adapts itself intelligently with environmental changes to learn the channel qualities dynamically [34]. In particular, in these methods, each network node acts as a learning agent which gradually estimates channel conditions based on the history of its experience with the channels in the past, and becomes more inclined towards channels with higher estimated success rate in its future transmissions. In general, model-free schemes are particularly interesting in most practical scenarios in which the model of the channel variations change over time, or it is not possible to obtain reliable statistical information about the process before node deployments. In the sequel, we review the related work on adaptive channel hopping in IEEE 802.15.4.

Table I
Used Main Acronyms

| | |
|---|---|
| **TSCH** | Time-Slotted Channel Hopping |
| **MAB** | Multi-Armed Bandits |
| **MABB** | Multi-Armed Bernoulli Bandits |
| **LoRa** | Long Range |
| **LoRaWAN** | Long Range Wide Area Network |
| **ASN** | Absolute Sequence Number |
| **AFF** | Adaptive Forgetting Factor |
| **PDR** | Packet Delivery Ratio |
| **ACK** | Acknowledgement Message |
| **TS** | Thompson Sampling |

### C. Related Work

The A-TSCH algorithm in [15] is among the pioneer work in adaptive channel hopping which provides enhanced reliability on the basis of the TSCH technique of IEEE 802.15.4-2015 standard. A-TSCH hops selectively among a subset of channels considered “reliable”, unlike TSCH which indiscriminately uses all 16 channels in the 2.4 GHz band. However, the evaluation of the channel quality in [15] is based on a costly spectrum sensing technique which requires detecting the ambient energy level to gauge the intensity of channel utilization. This process exerts a massive load on the system which eventually causes higher energy consumption and delay.

Alternatively, the ETSCH algorithm in [16] applies a non-intrusive channel quality estimation by energy detection during idle periods of time-slots. Then, the hopping sequence is propagated in the network including channels with high quality as whitelist. ETSCH is most efficient when the time-slot duration is sufficient to compensate clock drifts, while leaving enough time for energy detection. Handling multi-hop topologies remains a challenging issue and dedicated time-slots for energy detection may be required. In [36], the authors have presented an improved version of ETSCH with a Distributed

Channel Sensing mechanism (ETSCH+DCS) which finds channels with suitable conditions for packet transmission. A hybrid technique is employed for channel quality estimation which combines a central method with a distributed method. The central method uses Non-Intrusive Channel-quality Estimation (NICE) to compute energy consumption at the coordinator node, while the distributed channel estimation determines the sources of interference throughout the entire network. However, much overhead is still imposed on the end-nodes to identify interference sources hidden from the coordinator. Also, ETSCH+DCS assumes that the coordinator can directly communicate with all network nodes, which is not realistic in most industrial deployments.

In [20], a transmitter-side channel selection scheme is formulated as an independent process using packet transmission status (packet acknowledgement status) and Clear Channel Assessment (CCA) failures on that channel. The channel list is sent to the coordinator by augmenting it to the information element of the TSCH packet, and the coordinator broadcasts the newly updated list of channels. However, some implementation details are missing; for example, how the list of "good channels" is chosen. Further, it is required that the receiving node remain aware of the transmitted packet rate, which is not a realistic assumption in some event-driven sensing scenarios.

An approach for blacklist formation in TSCH-based networks is presented in [39]. In this method, each pair of nodes communicating with each other have their own blacklist. In particular, the nodes at each connection contact each other by exchanging control packets and agree on a blacklist. These control packets are sent at certain time-slots and use the ratio of received packets as a criterion to build the blacklist. However, non-blacklisted channels are selected at random without caring for scheduling constraints, which could cause interference between communication links.

In MABO-TSCH algorithm of [21], every pair of nodes locally blacklists physical channels. In order to reach an agreement between the transmitter and the receiver on the same blacklist, the list is interchanged via ACK packets to avoid network overhead. Using a MAB-based formalism, each physical channel is considered as an arm and the nodes estimate their PDR. Simulation results show that MABO-TSCH results in 23% more throughput compared to the basic frequency hopping technique, and that MAB is able to choose the best channels in 75% of cases. This method is particularly suited for networks with low neighborhood cardinality because negotiation for blacklist creation would involve too many connections between the neighbors.

In [17], every node holds a list of channels with unfavorable condition as well as channels that affect the connection negatively but are not yet placed permanently in blacklist. Once a channel is added to the temporary list, this information must be distributed in the entire network so that all the nodes hold the same blacklist for channel hopping. Also, when a channel is temporarily blacklisted, it will be ultimately blacklisted after a certain time. Some studies have conducted exhaustive experiments to showcase the importance of adaptive channel hopping as well as to evaluate the efficacy of the existing strategies. For example, in [37], an experiment has been conducted to evaluate the efficiency of a TSCH network in an airplane cabin affected by external disturbances generated by a Wi-Fi network. The investigation assumes there are 16 channels available and measures a packet error rate of approximately 35%. It is also concluded that throughput decrease whenever a smaller number of channels are used. In [35], the authors have extensively experimented with some reinforcement learning strategies for adaptive channel hopping. In the simulation process, PDR is employed to assess and compare the methods. External interference is simulated through placing two wireless connections in one location as well as multipath fading caused by reception of different replicas of a signal from diverse paths.

### D. Motivation

IEEE 802.15.4 uses the license-free 2.4 GHz band which is also exploited in other standards and protocols such as Wi-Fi; therefore, an efficient scheme is needed to combat the interference caused by co-existing entities. The default channel hopping in TSCH is not suitable for reducing the effect of interference because it is based on a circulatory simple formula that lacks the intelligence to enable proper selection of desirable frequency channels for hopping. Recently, some adaptive channel hopping methods have been published which attempt to learn the channels' conditions and create a blacklist of undesirable channels. To the best of our knowledge, previous studies in adaptive channel hopping for TSCH-based networks have all assumed a stationary statistical model for the external interference. However, in real world environments, it is reasonable to assume that the external interference, such as that caused by Wi-Fi or Bluetooth, often displays non-stationary behavior since the interfering sources often appear/disappear in a time-varying way and they also typically use the channels based on a dynamic policy that is not constant over time.

The non-stationarity of external interference has recently been addressed in designing new MAC mechanisms in LoRaWAN-based IoT settings [40 and 41]. For example, in [41], the authors have proposed two learning-based methods, (i.e., UCB and Thomson Sampling) for spectrum access in non-stationary IOT environments. However, these learning procedures have so many algorithmic parameters which should be tuned across a large number of IoT devices for proper execution. Also, LoRaWAN systems have some key differences with the case of IEEE 802.15.4 standard addressed in this paper: they operate in a different ISM frequency band (863-870 MHz in Europe and 902-928 MHz in US) which is not susceptible to 2.4 GHz Wi-Fi interference. This is while Wi-Fi constitutes the main source of interference for IEEE 802.15.4. Also, LoRaWAN is used for long-range single-hop node-to-gateway IoT communications, while IEEE 802.15.4 TSCH is typically used for multi-hop short range communications.

In this study, to handle non-stationary external interference in TSCH networks, we use the Dynamic MABB theory [23], and propose a lightweight algorithm to be applied iteratively by the IoT nodes to determine their hopping pattern dynamically.

## III. System Model

We consider a tree-like topology consisting of half-duplex wireless nodes similar to Fig. 1 in which each node sends data to its upstream (parent) node. The blue lines depict useful communication links, which extend from each child to its corresponding parent. The orange lines represent interference as we elaborate in the sequel. As can be seen from Fig. 1, there is a gateway node in which the data is gathered eventually.

### A. Interference Model

TSCH-based networks are susceptible to two types of interference: internal and external. Internal interference is due to the limitations imposed by the network topology and should be avoided through scheduling algorithms. The orange lines in Fig. 1 represent the internal interference. For example, a transmission from node 3 is not only heard by node 6, but it also affects node 7 as interference. On the other hand, external interference is due to the simultaneous use of the frequency band of the IEEE 802.15.4 by equipment that are based on other standards and protocols. The main focus of the present study will be on external interference. Our interference model is based on collision, which means that if interference occurs, there will be no weakening, but the useful signal will totally be corrupted [42].

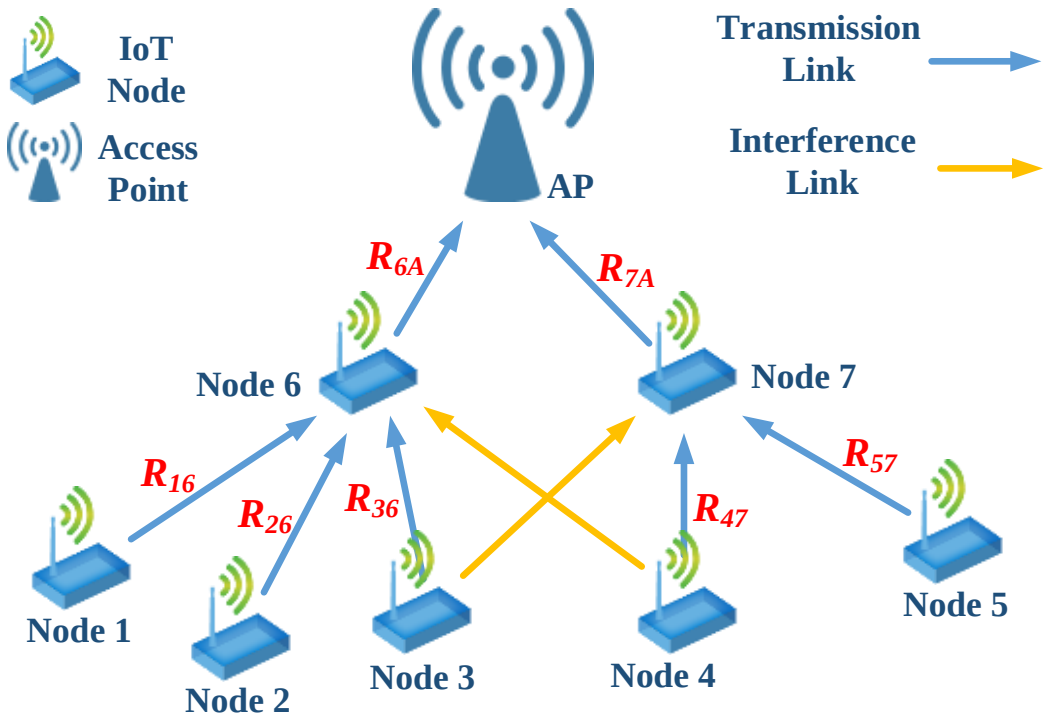

Fig. 1. An IEEE 802.15.4 TSCH network with a tree-like topology.

#### 1) *Stationary external interference*

In order to provide a mathematical definition for the stochastic process of presence or absence of the external interferer in IEEE 802.15.4 channels, we use the symbol $\mathbb{J} = \{1.\dots.j.\dots.J\}$ to denote the set of offsets of the frequency channel used in the TSCH schedule, and the symbol $\mathbb{I} = \{1.\dots.i.\dots.I\}$ to show the set of offsets of the frequency channels used by the interferer. We also represent the binary random variable of occurrence or absence of interference between an offset channel such as $i \in \mathbb{I}$ (of the interferer) and some offset channel such as $j \in \mathbb{J}$ (of IEEE 802.15.4) by the symbol $\gamma_{ij} \in \{0,1\}$, which has the probability mass function of $\mathbb{P}(\gamma_{ij} = 1) = p_{ij}$ and $\mathbb{P}(\gamma_{ij} = 0) = 1 - p_{ij}$, respectively.

Let the binary random variable $Y_j(n)$ indicate the occurrence or absence of interference in the $j^{th}$ offset during time-slot $n$, then the following assumption can be made on the stationary nature of the external interference:

**Assumption 1** (*stationary external interference*): The stochastic process $\{Y_j(n)\}_{n\in\mathbb{N}}$ is a stationary time series, when:

$$Y_j(n) = \gamma_j, \ \forall n \in \mathbb{N}, \forall j \in \mathbb{J} \quad (2)$$

where $\gamma_j \in \{0,1\}$ is a Bernoulli random variable with the following probability mass function:

$$\mathbb{P}(\gamma_j = 1) = p_j, \quad (3)$$

$$\mathbb{P}(\gamma_j = 0) = 1 - p_j, \quad (4)$$

$$p_j = \sum_{i\in\mathbb{I}} p_{ij}. \quad (5)$$

■

In other words, under the assumption of stationarity, the probability of interference occurrence for each channel offset remains constant over time.

#### 2) *Non-stationary external interference*

To describe precisely the behavior of the non-stationary interference, we use $\Psi_j$ to denote a finite and countable set of probabilistic regimes for the occurrence of interference on the $j$-th channel offset. Each probabilistic regime such as $\psi_j \in \Psi_j$ corresponds to a separate probability distribution such as $F_{\psi_j}(.)$, a finite average value of $\mu_{\psi_j}$, and a finite variance denoted by the symbol $\delta^2_{\psi_j}$. The symbol $T_{\psi_j}$ represents the random time during which regime $\psi_j$ governs the interference process. Also, let $\psi_j(n) \in \Psi_j$ represent the probabilistic regime that holds during time-slot $n$. Accordingly, we use $F_{\psi_j(n)}(.)$ to represent the probabilistic distribution corresponding to the current regime, $\delta^2_{\psi_j(n)}$ to represent the variance of this regime, and $\mu_{\psi_k(n)}$ represents its mean value.

Note that we do not make any assumption about the specifics of the regime switching process (i.e., replacement of the probabilistic regimes over time) as the approach presented in this paper is applicable to any general non-stationary interference pattern. That being said, however, for the purpose of simulation, we intentionally emphasize on a special regime switching process which occurs according to a first-order Markov chain (see Definition 1 below):

**Definition 1** (*Markovian switching non-stationarity*): The stochastic process $\{Y_j(n)\}_{n\in\mathbb{N}}$ is a non-stationary time series of with Markovian switching nature when $\psi_j$ corresponds to the state space of a discrete-time Finite-State Markov Chain (FSMC), and the transitions between any two regimes $\psi_j, \psi_{j'} \in \Psi_j$ occurs according to a stochastic matrix such as $Q = [q_{jj'}]$. ■

### B. Framing Model and Multi-Offset Scheduling

In compliance with IEEE 802.15.4 TSCH, the network-wide transmission schedule is represented in the form of a two-dimensional channel-slot matrix similar to Fig. 2, which repeats over time. A TSCH schedule determines which pair of nodes should exchange data packets on which channel and in which time-slot [32]. According to Eq. (1), although each link (or sender-receiver pair) has a constant position in the schedule, they will all eventually be assigned to different physical channel numbers over time (due to the cyclic pattern of the slot-frames).

Most existing algorithms for constructing TSCH schedules are single-offset in the sense that if a link is scheduled in a given time-slot, the transmission has to occur on the uniquely specified channel corresponding to that offset. This is while an adaptive channel hopping scheme needs to be built on a multi-offset schedule in which several offsets can be offered to each node at each time-slot. Realizing this, here we use a multi-offset variant of our proposed throughput-centric scheduling scheme in [31]. Our algorithm is centrally executed and has two phases: in the first phase, a conflict graph is formed that captures non-allowable simultaneous transmissions (e.g., due to single-radio half-duplex restrictions). We use graph-theoretical algorithms for the computation of independent sets in this conflict graph. Let $k$ and $\acute{k}$ be the two ends of a given communication link $\acute{k} \rightarrow k$. The output from phase one is the allocated slot-frame time-slots $\mathcal{T}_{\acute{k}\rightarrow k}$ for each link $\acute{k} \rightarrow k$. In the second phase, we aim at omitting internal interferences by accounting for hidden terminals. In each time-slot, distinct channel offsets are assigned to interfering links, while non-interfering links can be grouped as a single set and be associated with the same offset. Now, in each given time-slot, if unused offsets are still available, they will exhaustively be allocated to the links scheduled in that slot in such a way that the total network throughput is maximized. This way, if a link $\acute{k} \rightarrow k$ is scheduled for activation in a given time-slot, the output from phase two determines all the legitimate channel offsets $\mathcal{C}^{\mathcal{T}}_{\acute{k}\rightarrow k}$ from which a single channel can be chosen by our proposed hopping algorithm in Section V.C for actual transmission. More specifics on the computation of $\mathcal{T}_{\acute{k}\rightarrow k}$ and $\mathcal{C}^{\mathcal{T}}_{\acute{k}\rightarrow k}$ calls for a complete discussion of the scheduling algorithm, which remains outside the scope of this paper. As a final note, while the adopted scheduling algorithm is centralized, but the proposed channel hopping scheme is orthogonal to scheduling as long as it offers multiple offsets for each communication link in each time slot. In fact, the only interaction of our DMABB-CH scheme (Algorithm 1) with the underlying scheduling algorithm is in its input stage where the running node needs to receive the list of allocated slot-frame time-slots $\mathcal{T}_{\acute{k}\rightarrow k}$ and associated channel offsets $\mathcal{C}^{\mathcal{T}}_{\acute{k}\rightarrow k}$. Hence, DMABB-CH can as easily run alongside a distributed scheduling scheme as it can run over a centralized scheduling algorithm. As it turns out, the newly emerging distributed TSCH scheduling algorithms (such as ALICE in [47] and OST in [48]) work in a link-based fashion in the sense that they allocate cells to each directional link (a pair of nodes and traffic direction), and use multiple channels for the same time-slot.

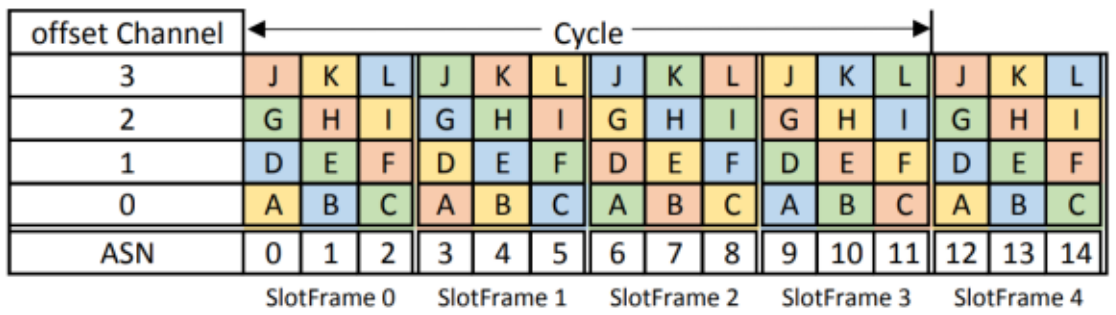

Fig. 2. Cycle structure in TSCH [42]

## IV. Problem Formalization as a Dynamic MABB

### A. Background on multi-armed bandits

The MAB problem is a classical optimization problem that explores the trade-off between exploitation and exploration in reinforcement learning. The problem consists of a machine with $J$ arms, and an agent that selects and pulls a sequence of arms, each of which generates some reward or penalty as a return for the agent. The goal of the agent is to minimize the *regret*, which is the difference between the reward gained by following a specific policy and the reward gained by selecting the best arm (in hindsight) after a sequence of arm selections. In MAB, the agent should compromise between exploration (discovering the unknown by selecting new arms to find their reward probability) and exploitation (using the formerly known arms to accrue high reward values). In standard MAB, it is assumed that the probabilistic model of rewards does not change over time, i.e., the optimal arm is the same in all time. A MAB problem with static reward distributions is also known as the stationary or static MAB problem in the literature (e.g., [43]).

The channel hopping problem can be readily mapped into a MAB setting in which each IoT device acts as the learning agent, the channels are considered as arms, and the expected PDR is defined to be the optimization objective. However, due to the time-varying nature of external interference (c.f., Section III.A.2), the optimal choice may change over time, and thus the assumption of static reward distributions is not adequate. As such, we should resort to non-stationary variants of the MAB formalism [33] where the underlying distribution of rewards for each arm may be time-varying (much the same as the probability of presence/absence of external interferences). Also, given that the outcome of a packet transmission (success/failure) can be described by a Bernoulli random variable, among the existing dynamic MAB formalisms, we adopt the one specifically tailored for dynamic "Bernoulli" processes, namely the "Dynamic MABB" [22], [23]. In this MAB formalism, the sequence of rewards/penalties obtained from each arm forms a timey-varying Bernoulli process with an unknown reward/penalty probability.

### B. Problem formulation

In order to formalize the channel hopping problem as a Dynamic MABB, we use $n \in \mathbb{N}$ to index discrete time. At each scheduled time-slot, the receiving node $k$ of each link $\acute{k} \rightarrow k$ chooses a channel $j(n)$ from the set of available channels $\mathbb{J}$.

**Remark 1**. In order to streamline notation, we drop the subscript $k$ in $j(n)$ and in almost all variables maintained internally by any decision making node $k$. Also, in Algorithm 1, we present a so-called DMABB-CH algorithm which can be executed independently by any given receiving node in the network topology. As such, the particular node index is mostly irrelevant to the discussion in the sequel. Where needed, however, the symbol $k$ has been used for more clarity. ■

The binary random variable $Y_j(n)$ determines the success or failure of the transmission over channel $j$ at time $n$. Let $\mu_j(n) = \mathbb{E}[Y_j(n)]$ to denote the mean reward for channel $j$ at time $n$. Our aim is to optimize the channel selection sequence and maximize the total expected reward (PDR) $\sum_{n=1}^{N} \mu_{j(n)}(n)$, or equivalently, minimize the total regret, defined as [22]:

$$\mathcal{R}(N) \stackrel{\text{def}}{=} \sum_{n=1}^{N} \mu_{j^*(n)}(n) - \sum_{n=1}^{N} \mu_{j(n)}(n), \qquad (6)$$

$j^*(n)$ is the optimal channel at time $n$ and can be described as:

$$j^*(n) = arg \max_{j \in \mathbb{J}} \mu_j(n) \quad (7)$$

Intuitively, we would like the regret to be as small as possible. In Section VI, we propose a learning-based algorithm that is capable of achieving and maintaining a sub-linear regret value by swiftly identifying the switching of the optimal arm. In fact, a sub-linear regret value signifies that the expected difference in total rewards obtained by an optimal policy (the policy that chooses $J^*(n)$ in every moment) and the total reward actually earned by the device vanishes in the long run.

## V. The Proposed Adaptive Hopping Algorithm

Several popular procedures exist for arm selection in static MAB problems, namely; $\varepsilon$-Greedy [24], Upper Confidence Bound (UCB) [25], and Thompson Sampling (TS) [26]. These procedures start with exploration and as the process goes forward, and experience accumulate, switch to exploitation, i.e., *converging* to only selecting the optimal arm, simply by selecting the arms with frequencies proportional to their probabilities of being optimal. These techniques are designed for settings where the reward probabilities of the bandit arms remain constant. In non-stationary scenarios, where the reward probabilities are dynamically evolving, we need procedures that can *track* the potential reward probability changes.

In a Dynamic MABB setting, an estimator needs to be in place to track the expectation of rewards via putting more weight on the more recent reward history. Some example procedures include: the $\varepsilon$-Greedy algorithm coupled with an exponentially-weighted moving average estimator [28], sliding-window UCB (SW-UCB) [29], and Dynamic Thompson Sampling (DTS) [30]. It is worth noting that, as argued in [22], these procedures are based on complex algorithms that call for precise tuning of some of their configuration parameters, which are dependent on the knowledge of the underlying stochastic process. For instance, to tune the window size of SW-UCB, the number of switch points must be known (i.e., how many times the optimal arm switches).

Lately, a number of innovative schemes have been reported in [22] that are specifically crafted for computing a learning policy in dynamic MABB problems. These schemes are quite easy to implement, and far less complicated compared to prior procedures in the sense that the tuning of key parameters is automated, and there is no need for prior knowledge of the model structure. These features are particularly useful for applicability in IoT devices which operate on low-quality hardware (e.g., small embedded processors) and offer limited software and power capabilities (i.e., very small -size batteries).

Exploring the proposed schemes in [22], we choose the AFF-OTS procedure as the basis for designing our channel hopping algorithm. Actually, AFF-OTS involves fewer computational steps, allows for more flexible tuning and, as demonstrated through exhaustive simulations in [22], outperforms the competing schemes in terms of total regret.

The learning-theoretic framework presented in [22], describes a two-step procedure for tracking the optimal arm in a dynamic MABB problem: *estimation* step and *selection* step. In our context, these steps would correspond to learning the reward distribution of each channel and selecting one channel to transmit, respectively. Accordingly, in the following section, we provide a short description of how channel hopping can take place adaptively using AFF-OTS as a dynamic MABB solver. To prevent redundancy while ensuring that the paper remains self-contained, we only present the key ideas, leaving out much of the technicalities, and refer the interested reader to [22].

### A. Estimating the expected PDR of each channel

In order to make a proper channel selection at each time step, each node must correctly and efficiently trace the expected PDR of the channels, especially when it is affected by time-evolving interference. To this end, the AFF-OTS algorithm resorts to an adaptive estimation technique in which the weight on older feedbacks is gradually decreased as new feedbacks come in. More specifically, an adaptive forgetting factor $\lambda$ (AFF for short) is applied whose value can be adjusted at each time step to ensure improved adaptation. Such an estimator would be capable of responding swiftly to changes in external interference without having prior knowledge of this process.

Assume a single channel, and suppose that $\{Y_n\}_{n=1:N}$ is the history of successful/unsuccessful transmissions over this channel up to time $N$. Eq. (8) gives a sample mean $\hat{Y}_N$ of the sequence $\{Y_n\}_{n=1:N}$ in an "adaptive forgetting" manner:

$$\hat{Y}_N = \frac{1}{w_N} \sum_{\tau=1}^{N} \left( \prod_{p=\tau}^{N-1} \lambda_p \right) Y_\tau \quad (8)$$

in which:

$$w_N = \sum_{\tau=1}^{N} \left( \sum_{p=\tau}^{N-1} \lambda_p \right) \quad (9)$$

is a normalizing constant to make the estimate $\hat{Y}_N$ unbiased for the case when the underlying interference process is i.i.d. It is worth noting that a more efficient method to compute $\hat{Y}_N$ is to run incremental updates as listed in Eqs. (10) to (12):

$$\hat{Y}_N = \frac{m_N}{w_N} \quad (10)$$

$$m_N = \lambda_{N-1} m_{N-1} + Y_N \quad (11)$$

$$w_N = \lambda_{N-1} w_{N-1} + 1 \quad (12)$$

The AFF sequence $\boldsymbol{\lambda}_N \stackrel{\text{def}}{=} \{\lambda_n\}_{n=1:N}$ applied to the above equations should produce a desirable tracking performance (e.g., in terms of the one-step-ahead squared prediction error $L_n \stackrel{\text{def}}{=} \left(\hat{Y}_{n-1} - Y_n\right)^2$ between the sample mean $\hat{Y}_{n-1}$ and the latest observed reward $Y_n$); in particular, AFF $\lambda_n$ is computed via a single gradient descent step:

$$\lambda_n = \lambda_{n-1} - \eta \Delta(L_n, \lambda_{n-2}) \quad (13)$$

where $\eta (\eta \ll 1)$ is the step size, and $\Delta(L_n, \lambda_{n-2})$ is given as:

$$\Delta(L_n, \lambda_{n-2}) = \lim_{\epsilon \to 0} \frac{L_n(\lambda_{n-2}+\epsilon) - L_n(\lambda_{n-2})}{\epsilon} \quad (14)$$

where for any $\lambda$, $L_n(\lambda)$ means evaluating $L_n$ using the forgetting factor $\lambda$. Also, the expression in (14) may be interpreted as the derivative of $L_n$ w.r.t. the forgetting factors $\lambda_1,\dots, \lambda_{n-2}$. Therefore, $\Delta(L_n, \lambda_{n-2})$ is an online derivative-like function for $L_n$ w.r.t. $\lambda_{n-2}$ which contains pre-computed values; if the value of $\lambda_n$ which was calculated using (13) is greater than 1 (or less than 0), it is truncated to 1 (or 0) to ensure that $\lambda_n \in [0,1]$.

Finally, the following recursions is required in order to sequentially compute $\Delta(L_n, \lambda_{n-2})$:

$$\Delta(L_n,\lambda_{n-2}) = 2(\hat{Y}_{n-1} - Y_n)\left(\frac{\dot{m}_{n-1}-\dot{w}_{n-1}\hat{Y}_{n-1}}{w_{n-1}}\right) \tag{15}$$

$$\dot{m}_n = \lambda_{n-1}\dot{m}_{n-1} + m_{n-1} \tag{16}$$

$$\dot{w}_n = \lambda_{n-1}\dot{w}_{n-1} + w_{n-1} \tag{17}$$

where $\dot{m}_1 = 0$, and $\dot{w}_1 = 0$. Note that $\dot{m}_{n-1}$ and $\dot{w}_{n-1}$ are defined as the derivative of $m_{n-1}$ and $w_{n-1}$ , respectively, and w.r.t. $\lambda_{n-2}$, which are similar to $\Delta(L_n, \lambda_{n-2})$. However, they are denoted by overhead dot rather than $\Delta(m_{n-1}, \lambda_{n-2})$and $\Delta(w_{n-1}, \lambda_{n-2})$ for notational simplicity.

TABLE II
SUMMARY OF SYMBOLS

| Symbol | Description |
|---|---|
| $j \in \mathbb{J}$ | frequency channel offset in IEEE 802.15.4 (TSCH) |
| $i \in \mathbb{I}$ | frequency channel offset of external interferers (Wi-Fi/BLE) |
| $SF_{length}$ | slot-frame length |
| $P_{ij}$ | interference probability between the $i$ and $j$ channels |
| $\gamma_{ij}$ | binary random variable of occurrence or absence of interference between the $i$ and $j$ channels |
| $\mathbb{P}$ | probability mass function of $\gamma$ |
| $\Psi_j$ | probabilistic regimes for the interference on the $j$ channel |
| $n$ | time index |
| $F_{\psi_j}$ | probability distribution of the regime $\psi_j$ |
| $\mu_{\psi_j}$ | finite average value of $F_{\psi_j}$ |
| $T_{\psi_j}$ | random duration of establishing the regime $\psi_j$ |
| $\delta^2_{\psi_j(n)}$ | to represent the variance of the regime $\psi_j$ |
| $Z_j(n)$ | stochastic process of the switching nature |
| $Q$ | regimes transitions matrix |
| $\lambda$ | adaptive forgetting factor (AFF) |
| $\{Y_n\}_{n=1:N}$ | history of rewards gained from a channel up to time $N$ |
| $\hat{Y}_N$ | sample mean of the sequence $\{Y_n\}_{n=1:N}$ |
| $w_N$ | normalizing constant |
| $L_n$ | one-step-ahead squared prediction error |
| $\eta$ | step size |
| $\Delta$ | derivative function for $L_n$ |
| $Beta()$ | conjugate prior to the Bernoulli distribution |
| $\alpha_n$ | the first *Beta* distribution parameter at time $n$ |
| $\beta_n$ | the second *Beta* distribution parameter at time $n$ |

### B. Updating PDR estimates for unselected channels

Each IoT node in our proposed setting will keep a PDR estimate for each TSCH channel. However, a given node can only observe one channel at a time which means that the estimations and intermediate quantities of an unobserved channel will retain their previous values (if channel $j$ is not observed at time $n$). Not being able to update estimators gives rise to more challenges in dynamic cases. Although the estimator tracks the expected reward thoroughly at any given instant, the tracking precision may decrease quickly once it stops taking new observations, making it harder to achieve a balance between exploration and exploitation. We can make changes to the procedure that was presented in Section VI.A to keep updating PDR estimates even for unselected channels by discounting $m_n$ and $w_n$. In particular, we introduce new quantities $\widetilde{m}_n$ and $\widetilde{w}_n$ which are computed as:

$$\widetilde{m}_n = (\lambda_n)^{\frac{n-n_{last}}{|J|}} m_n \tag{18}$$

$$\widetilde{w}_n = (\lambda_n)^{\frac{n-n_{last}}{|J|}} w_n \tag{19}$$

where $n_{last}$ denotes the last time that the channel was selected. Note that if a channel is actually chosen by an IoT node, the two sets of quantities are identical; i.e., $\widetilde{m}_n = m_n$ and $\widetilde{w}_n = w_n$. On the contrary, when a channel is not selected, $m_n$ and $w_n$ are discounted by the forgetting factor that was obtained when it was selected the last time (bear in mind that) the forgetting factor $\lambda_n$ of an unselected arm remains the same as $\lambda_{n_{last}}$.

### C. Adaptive channel selection scheme

Now that we have seen how an IoT node can estimate and track the mean PDR associated with each channel, we can describe the channel selection process based on the AFF-OTS algorithm in [22]. AFF-OTS is itself based on standard Thompson Sampling (TS) [26] procedure in MAB theory. The idea in TS is to form and update Bayesian beliefs on the expected reward of each arm. In particular, an initial belief (a so-called "conjugate prior") is first assigned to the expected reward of each channel, and then the posterior distribution of the expected reward is incrementally updated through successive channel selections. Now, a decision rule is constructed based on this posterior distribution: At each round, a random sample is drawn from the posterior distribution of each channel, and the channel with the highest sample value is selected. Now, in our Bernoulli bandit problem, we have to select the *Beta* distribution, $Beta(\alpha_0, \beta_0)$, as the standard conjugate prior to the Bernoulli distribution. The posterior distribution is then $Beta(\alpha_n, \beta_n)$ at time $n$, and the parameters $\alpha_n$ and $\beta_n$ can be updated recursively as follows:

$$\alpha_n = \alpha_0 + \widetilde{m}_n \tag{20}$$

$$\beta_n = \beta_0 + \widetilde{w}_n - \widetilde{m}_n \tag{21}$$

In other words, by using the discounted quantities $\widetilde{m}_n$ and $\widetilde{w}_n$, exploration of unselected channels are boosted (reinforced). More specifically, the posterior distribution is flattened for an unselected channel, and the longer the channel is unselected, the further its posterior distribution is flattened. The pseudo-code for the complete algorithm is given in Algorithm 1 (which is referred to as DMABB-CH). Also, important notations are collected in Table II for easier reference. In the beginning, a very short initialization period is applied to make initial estimations. Normally, the initialization duration is $O(|J|)$, i.e., selecting each channel only once. In line 8, the drawn sample value $x(j)$ for channel $j$ is replaced by its posterior mean $\zeta$ provided that the latter score is greater. This means that for each channel the score on which a decision is made will never be smaller than the posterior mean. This is because the AFF-OTS algorithm is based on a somewhat modified version of the conventional TS procedure, which is known as Optimistic Thompson Sampling (OTS) [44]. OTS leads to added improvement regarding exploration of highly uncertain channels compared to TS and the reason as it increases the probability of attaining a high score for channels with greater posterior variance.

Now, we discuss how DMABB-CH is actually executed in the network. As with most tree-based data collection applications, we also consider "receiver-based" channel selection [21]. Hence, every receiver node indexed by $k$ (the parent in link $\acute{k} \rightarrow k$) executes the algorithm independently for every link it has with its immediate child nodes. In each scheduled time-slot $n \in \mathcal{T}_k$, node $k$ tunes on its previously channel $j_n \in \mathcal{C}_k^{\mathcal{T}}$ to receive data from $\acute{k}$. In case of a successful reception, the Bernoulli reward $Y_n(j_n)$ is set to 1 and 0 otherwise. Also, node $k$ selects the channel $j_{n+1}$ for the upcoming transmission opportunity and notifies the corresponding sender $\acute{k}$ by piggybacking the chosen channel index on the feedback message for the recent transmission; i.e., either as part of the Ack for a successfully received message or along with the Nack for an expected (yet missed) message. This piggybacking causes negligible overhead.

**Algorithm 1** Pseudo-code of DMABB-CH algorithm running on a receiver node $k$ for each communication link $\acute{k} \rightarrow k, \forall \acute{k} \in child(k)$

**input:** Allocated time-slots $\mathcal{T}_{\acute{k}\rightarrow k}$ and associated channel offsets $\mathcal{C}_{\acute{k}\rightarrow k}^{\mathcal{T}}$
**begin**
*// INITIA1LIZATION*
Set initial shape parameters for Beta distribution: $\alpha_0(j), \beta_0(j)$ for $\forall j \in \mathbb{J}$;
Set $\eta \in (0, 1)$;
**for** $J = |\mathbb{J}|$ *slot-frames* **do**
  Select some fixed offset from $\mathcal{C}_{\acute{k}\rightarrow k}^{\mathcal{T}}$ and run default TSCH to receive over all physical channels $j \in \mathbb{J}$ according to the standard hopping sequence;
**end for**
*// MAIN*
00: **for** $n$=*ASN*, . . ., $N$ **do** // from current absolute sequence number onward…
01: **if** $n$ (mod *SF_length*) $\in \mathcal{T}_{\acute{k}\rightarrow k}$ **then** *// link $\acute{k} \rightarrow k$ is scheduled in time-slot n*
02: Tune to receive over channel $j_n$;
03: Determine reward $Y_n(j_n) = \begin{cases} 0, & no\ message\ received\ from\ \acute{k} \\ 1, & message\ received\ from\ \acute{k} \end{cases}$;
04: **for** all $j \in \mathbb{J}$ **do**
05: Update $\alpha_n(j)$ and $\beta_n(j)$ according to (20) and (21);
06: Draw a sample $x(j)$ from $Beta(\alpha_n(j), \beta_n(j))$;
07: **if** $x(j) < \zeta \overset{\text{def}}{=} \frac{\alpha_n(j)}{\alpha_n(j)+\beta_n(j)}$ **then**
08: Replace $x(j)$ with $\zeta$;
09: **end if**
10: **end for**
11: Select channel $j_{n+1} = \arg\max_{j' \in \mathbb{J}(\mathcal{C}_k)} x(j')$;
12: Send feedback (Ack/Nack) message to node $\acute{k}$ embedding $j_{n+1}$;
13: **end if**
14: **end for**
15: **end**

## D. Discussion on computational complexity

In this section, we discuss how much work needs to be performed by the algorithm in each iteration. In each iteration of the *MAIN* loop, the receiving node $k$ initially updates the Beta distribution parameters for all possible channels (typically 16 channels in IEEE 802.15.4) based on its recent reception experience on the chosen channel, i.e., $Y_n(j_n)$. In order to update the parameters $\alpha_n(j)$ and $\beta_n(j)$, we utilize the simple formulae in (20) and (21). These formulae entail primitive addition/subtraction operations over the new values for the parameters $\widetilde{m}_n$ and $\widetilde{w}_n$. Also, according to (18) and (19), $\widetilde{m}_n$ and $\widetilde{w}_n$ are in turn computed from the new values of $m_n$, $w_n$, and $\lambda_n$. However, these parameters are also updated using simple recursive/incremental equations from their previous values using a few primitive addition/subtraction/multiplication operations. Then, to determine the channel for the next transmission, node $k$ samples a value $x(j)$ from the Beta distribution for each channel $j$. This is just a basic random number generation. Next, a comparison is made between $x(j)$ and $\zeta$ to determine the final value for $x(j)$, and we exit the loop on the channel set. Following that, channel selection is made by finding $\arg\max_j x(j)$, to identify the channel with the largest $x(j)$. In sum, all the operations performed in each iteration of the DMABB-CH algorithm are as follows:

- Random number generation (once for each channel, 16 total)
- Simple comparison (once for each channel, 16 total)
- Maximization (over the channel set with 16 elements)
- Evaluation of some very simple recursive update equations (once for each channel, 16 total)

Each parent node has to execute this algorithm independently for each of its children. However, in each time-slot, only one instance of DMABB-CH is running on a given node, which has very low processing overhead. Also, assuming a tree topology with even a moderate branching factor, the memory footprint would also be reasonably small.

# VI. Simulation Results

## A. Simulation setup

### 1) Simulation Parameters

We set up a network with a tree topology deployed in a 200 × 200 m$^2$ area. Each node is positioned in a random place in the network and has a coverage radius of 50m and a neighborhood cardinality varying in the range [2, 20]. We evaluate the performance in several scenarios, varying the total number of nodes from 20 to 40. For evaluating the energy consumption, we use the measurements on the GINA mote presented in [45]. NS-3 [38] is used as simulator and the simulation parameters are specified as in Table III. We simulate Wi-Fi interference as the main culprit for external interference.

TABLE III
Parameters and Values used in Simulation

| Parameter | Value |
|---|---|
| Number of nodes | 30 (default, varies in some experiments) |
| IoT device TX power | 10 mW |
| Simulation duration | 20,000 slots |
| Packet size | 100 bytes |
| Network offered load (default) | 500 Kb/s |
| Number of channel (\|J\|) | 16 |
| Time-slot duration ($\tau_{slot}$) | 10 msec |
| Slot-frame length ($SF_{length}$) | 8 |

#### a) Stationary interference parameters

Under stationary interference, we use a single collision probability matrix to simulate the interference between each of the 802.11 and 802.15.4 channels, as can be seen in Table IV. Each row represents channel offsets in the 802.15.4 standard, with 16 offsets specified from 11 to 26, and each column denotes the 802.11g offsets. The number in each cell indicates the probability of two channels interfering.

*b) Non-Stationary interference parameters*

In the non-stationary mode, the statistical pattern of presence or absence of interference may vary over time. We simulate this statistical pattern using four interference probability matrices (similar to the sample matrix given in Table IV). Switching between these probabilistic regimes is generally governed by a Markovian process. More details are given in the sequel.

TABLE IV
INTERFERENCE PROBABILITY MATRIX IN STATIONARY MODE

| 802.11 channels / 802.15.4 channels | 1 | 2 | 3 | 4 | 5 | 6 | 7 | 8 | 9 | 10 | 11 | 12 | 13 |
|---|---|---|---|---|---|---|---|---|---|---|---|---|---|
| **11** | 0.44 | 0.03 | 0 | 0 | 0 | 0 | 0 | 0 | 0 | 0 | 0 | 0 | 0 |
| **12** | 0.6 | 0.35 | 0.04 | 0 | 0 | 0 | 0 | 0 | 0 | 0 | 0 | 0 | 0 |
| **13** | 0.33 | 0.7 | 0.53 | 0.04 | 0 | 0.02 | 0 | 0 | 0 | 0 | 0 | 0 | 0 |
| **14** | 0.08 | 0.42 | 0.74 | 0.35 | 0.05 | 0.08 | 0.02 | 0.01 | 0 | 0 | 0 | 0 | 0 |
| **15** | 0 | 0.1 | 0.49 | 0.82 | 0.24 | 0.12 | 0.04 | 0.03 | 0 | 0 | 0 | 0 | 0 |
| **16** | 0 | 0 | 0.06 | 0.44 | 0.78 | 0.28 | 0.11 | 0.07 | 0.01 | 0 | 0 | 0 | 0 |
| **17** | 0 | 0 | 0.01 | 0.13 | 0.33 | 0.69 | 0.34 | 0.1 | 0.03 | 0 | 0 | 0 | 0 |
| **18** | 0 | 0 | 0 | 0.07 | 0.11 | 0.28 | 0.74 | 0.42 | 0.15 | 0.02 | 0 | 0 | 0 |
| **19** | 0 | 0 | 0 | 0.01 | 0.06 | 0.15 | 0.28 | 0.85 | 0.48 | 0.16 | 0 | 0 | 0 |
| **20** | 0 | 0 | 0 | 0 | 0.03 | 0.07 | 0.07 | 0.65 | 0.78 | 0.51 | 0.08 | 0 | 0 |
| **21** | 0 | 0 | 0 | 0 | 0.01 | 0.03 | 0.02 | 0.04 | 0.66 | 0.79 | 0.37 | 0.06 | 0 |
| **22** | 0 | 0 | 0 | 0 | 0 | 0.01 | 0.02 | 0 | 0.07 | 0.54 | 0.88 | 0.41 | 0.03 |
| **23** | 0 | 0 | 0 | 0 | 0 | 0 | 0 | 0 | 0 | 0.09 | 0.4 | 0.69 | 0.48 |
| **24** | 0 | 0 | 0 | 0 | 0 | 0 | 0 | 0 | 0 | 0 | 0.03 | 0.47 | 0.83 |
| **25** | 0 | 0 | 0 | 0 | 0 | 0 | 0 | 0 | 0 | 0 | 0 | 0.06 | 0.48 |
| **26** | 0 | 0 | 0 | 0 | 0 | 0 | 0 | 0 | 0 | 0 | 0 | 0 | 0.09 |

2) *Compared methods*

In order to evaluate the performance of the proposed solution, comparison is made against two baseline schemes:

- **Default channel hopping in TSCH [7]**: There is no intelligence in the standard channel hopping of TSCH which uses the default non-adaptive periodic hopping formula (Eq. (1)). As such, a node may blindly use desirable/undesirable channels alike. In this scheme, it is also enough to use a single-offset scheduler.
- **MAB-based channel hopping [21]**: In this method, each node, using a Q-learning-based MAB algorithm, estimates the quality of the available physical channels, and at each round, selects the channel with highest estimated PDR. The method in [21] also uses a multi-offset schedule. It, however, follows a static MAB formulation which can only combat stationary 2.4 GHz interference.

3) *Scenarios*

We consider three main scenarios to simulate Wi-Fi interference: stationary interference, slow Markov non-stationary scenario, and fast Markov non-stationary scenario. We also consider a fourth practical interference pattern to account for stochastic interference processes with memory. However, due to space considerations, we conduct limited experiments with this fourth model only in terms of average PDR (c.f., Section VI.B.2). In both slow and fast Markov scenarios, the probabilistic regime of the interference can switch between four tables, i.e. Table IV and three other similar tables (not shown here to save space). In fact, each of these four probability matrices of interference is considered as a state of a Markov chain, and transitions from one regime to another happens according to the transition kernel given in Table V in the slow Markov case and according to Table VI in the fast Markov scenario. Unlike the fast case, a slow Markov transition matrix has a near-diagonal structure so as to elongate the expected duration time between jump changes that the system parameters remain constant. As for stochastic interference with memory, the interferer activity is modeled as a $k$th-order Markov chain, resulting in interference with memory up to $k$ time-slots. In the experiments, we only consider three memory levels (corresponding to first, second, and third-order Markov chains). In fact, a first-order Markov model has a 10 msec history, a second-order process has 20 msec, and a third order Markov model accounts for a 30 msec history. Knowing the transition probabilities for each, artificial traces are generated from all the three models and the algorithms are evaluated against these traces. The results have been averaged over 50 independent chain realizations.

TABLE V
PROBABILITY TRANSITION MATRIX FOR SLOW MARKOV SCENARIO

| | State 1 | State 2 | State 3 | State 4 |
|---|---|---|---|---|
| **State 1** | 0.85 | 0.05 | 0.05 | 0.05 |
| **State 2** | 0.05 | 0.85 | 0.05 | 0.05 |
| **State 3** | 0.05 | 0.05 | 0.85 | 0.05 |
| **State 4** | 0.05 | 0.05 | 0.05 | 0.85 |

TABLE VI
PROBABILITY TRANSITION MATRIX FOR FAST MARKOV SCENARIO

| | State 1 | State 2 | State 3 | State 4 |
|---|---|---|---|---|
| **State 1** | 0.1 | 0.4 | 0.35 | 0.15 |
| **State 2** | 0.25 | 0.05 | 0.3 | 0.4 |
| **State 3** | 0.25 | 0.3 | 0.05 | 0.4 |
| **State 4** | 0.4 | 0.25 | 0.3 | 0.05 |

## B. Simulation Results

In this section, we discuss the simulation results on the convergence properties of our algorithm, PDR, energy efficiency as well as the average end-to-end delay.

1) *Convergence*

In order to show the convergence behavior of our DMABB-CH algorithm, we only focus on the most challenging case which is the fast switching environment. A similar behavior can be seen in the other two cases but with a less noisy waveform and relatively faster convergence speed. The plot in Fig. 3 illustrates the normalized regret of the proposed learning policy in the fast Markov scenario as a function of time. As expected, the regret approaches zero as the time horizon grows large.

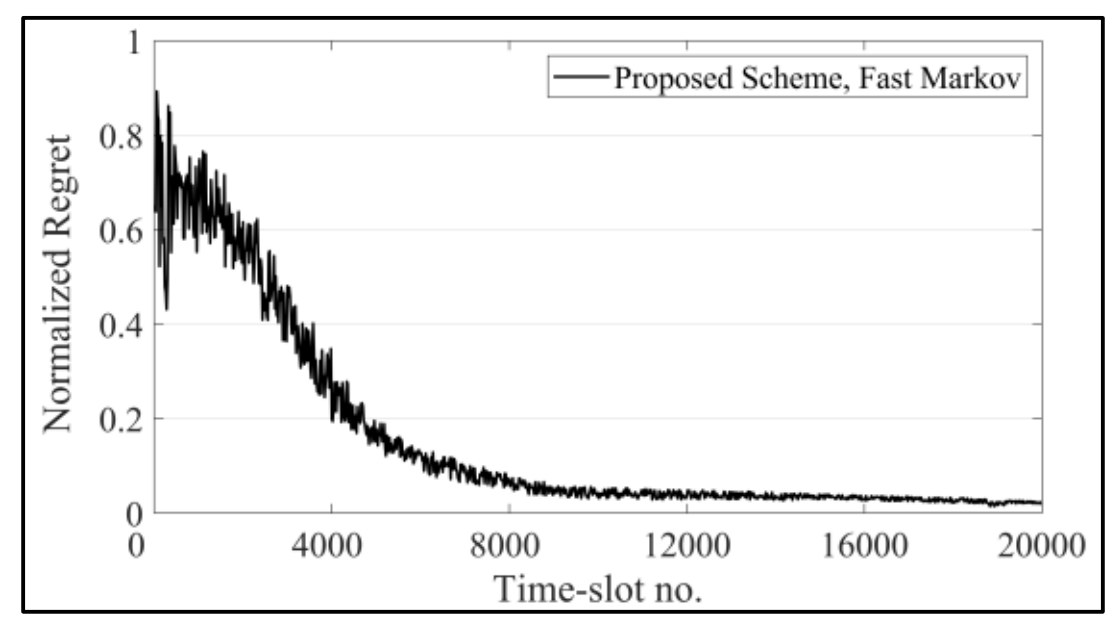

Fig. 3. Average regret over time

2) *Packet Delivery Ratio*

PDR is the ratio of the number of packets correctly received at the destination to the number of packets sent out by a source. Under the settings shown in Table III, Fig. 4 plots the average PDR for the three methods under all the three scenarios. As a general observation, the dynamics of the environment has a deteriorating impact on PDR performance. This impact is worst for the pure channel hopping of default TSCH as it has no intelligence for adaptation. Under a stationary pattern of interference, the MAB-based hopping in [21] has almost a similar limiting PDR as ours, but performs progressively worse under higher interference dynamics. In Fig. 5, we plot the average PDR as the network size increases in the number of nodes. For this experiment, all the simulation parameters are as shown in Table III except for the number of nodes which varies from 20 to 40. In general, if the number of scheduling resource blocks (cells) remains the same, PDR decreases in more crowded topologies. Fig. 5 shows this decreasing trend for the algorithms in all three configurations. Next, in Fig. 6, we plot the average PDR versus the variations in the offered load. In a topology of 30 nodes, we simulate both lightly and heavily-loaded scenarios by varying the load introduced by the active connections. Similarly, to the case of increasing network size, the PDR performance degrades as the load intensifies.

The results are given in Table VII. In general, in higher order models, a node running a MAB-based algorithm remembers more "history", and since additional history gives the node more predictive power, the performance of the learning algorithm has improved accordingly. In contrast, the PDR performance of the standard TSCH has no meaningful relationship with the memory level of the stochastic interference process.

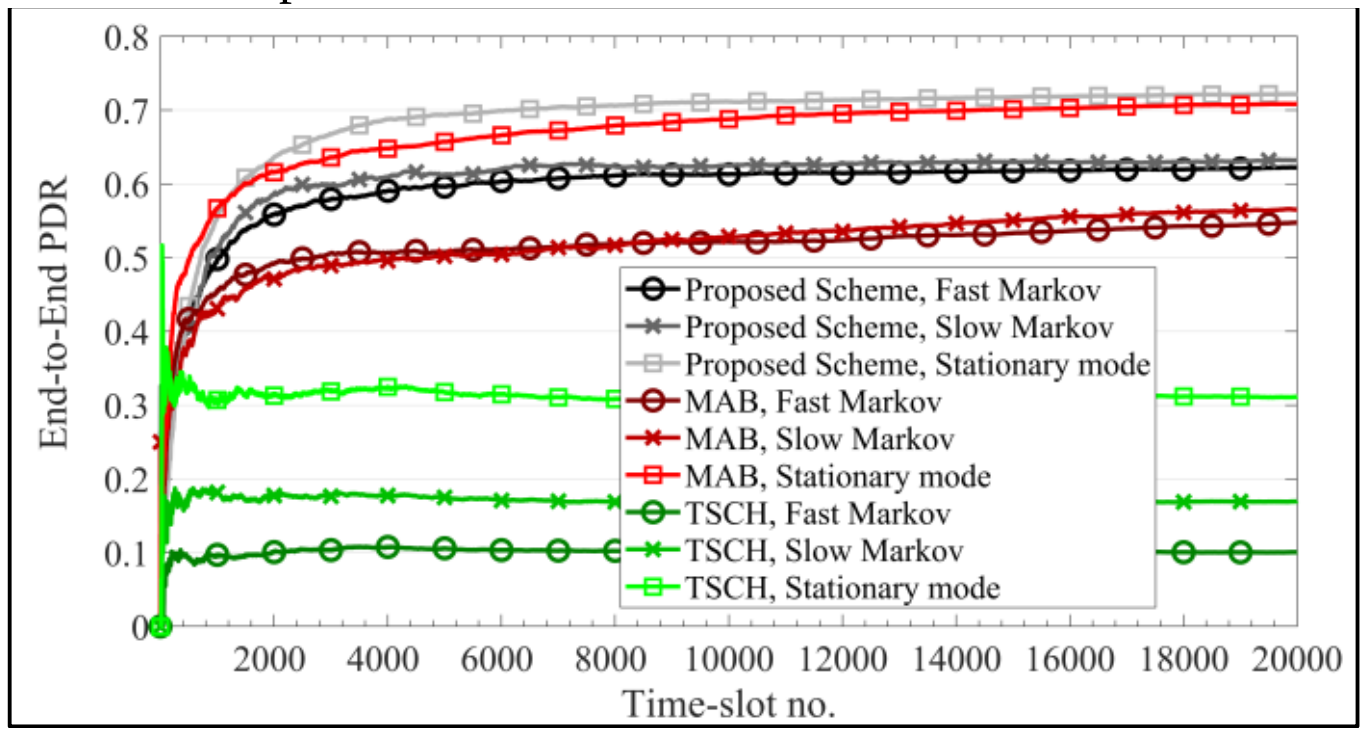

Fig. 4. Average PDR over time

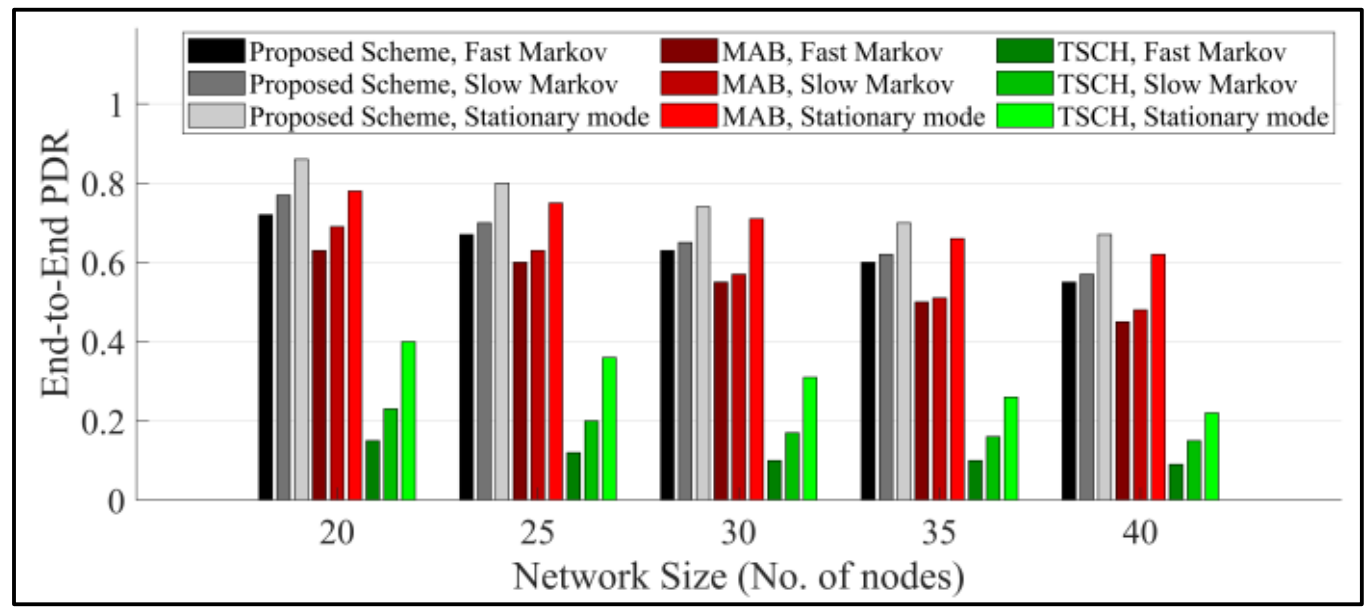

Fig. 5. PDR vs. Network size.

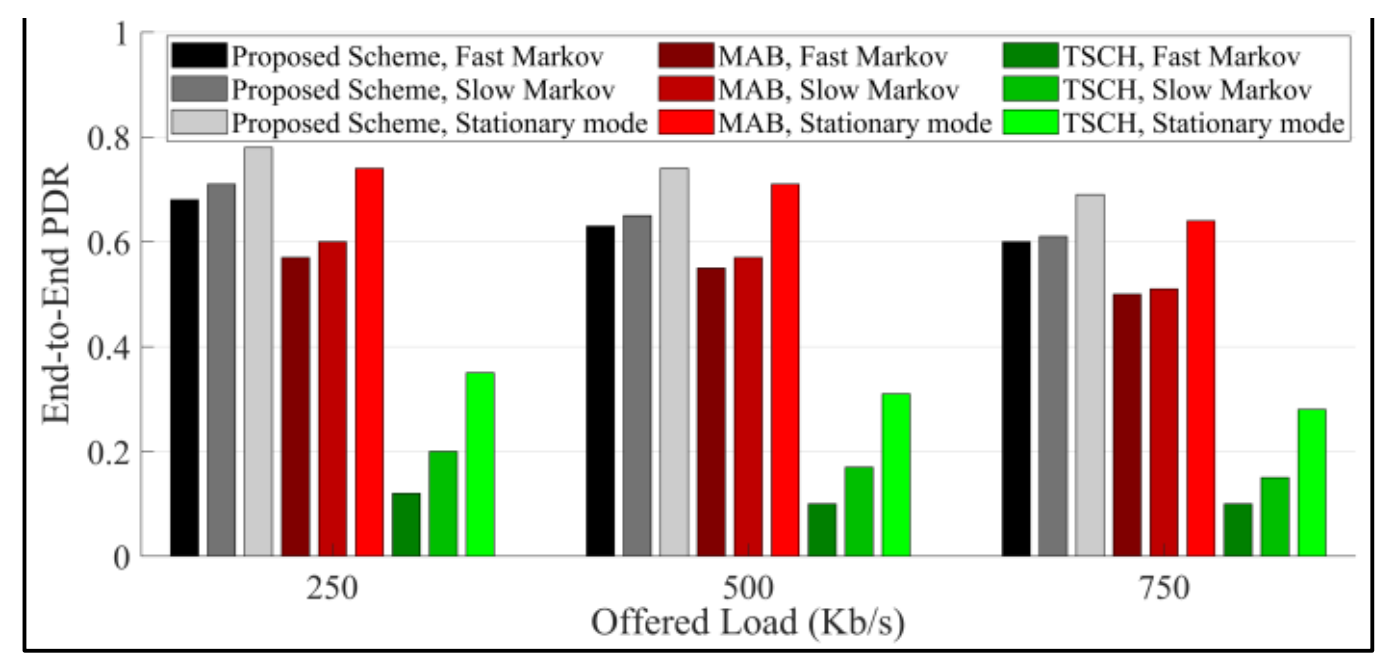

Fig. 6. PDR vs. Offered load.

TABLE VII
PDR UNDER INTERFERENCE PROCESS WITH MEMORY

| MEMORY LEVEL | PROPOSED (DMABB-CH) | MAB | TSCH |
|---|---|---|---|
| 1 | 0.69 | 0.64 | 0.28 |
| 2 | 0.75 | 0.71 | 0.31 |
| 3 | 0.82 | 0.76 | 0.29 |

### 3) *Energy efficiency*

In this experiment, we focus on energy efficiency defined as the energy required per successfully delivered packet (or bit). An intelligent and adaptive channel hopping strategy would be able to reduce the total number of unnecessary packet transmissions and receptions in the global network. The results in Fig. 7 were obtained for the duration of simulation in which packets flow from downstream nodes all the way up to the gateway. The plot sufficiently demonstrates that a lower number of packets are sent by each sensor node when the proposed approach is used when compared to the two baseline approaches. Under all network sizes, the proposed approach improves the overall energy efficiency of the network by reducing the number of retransmission attempts.

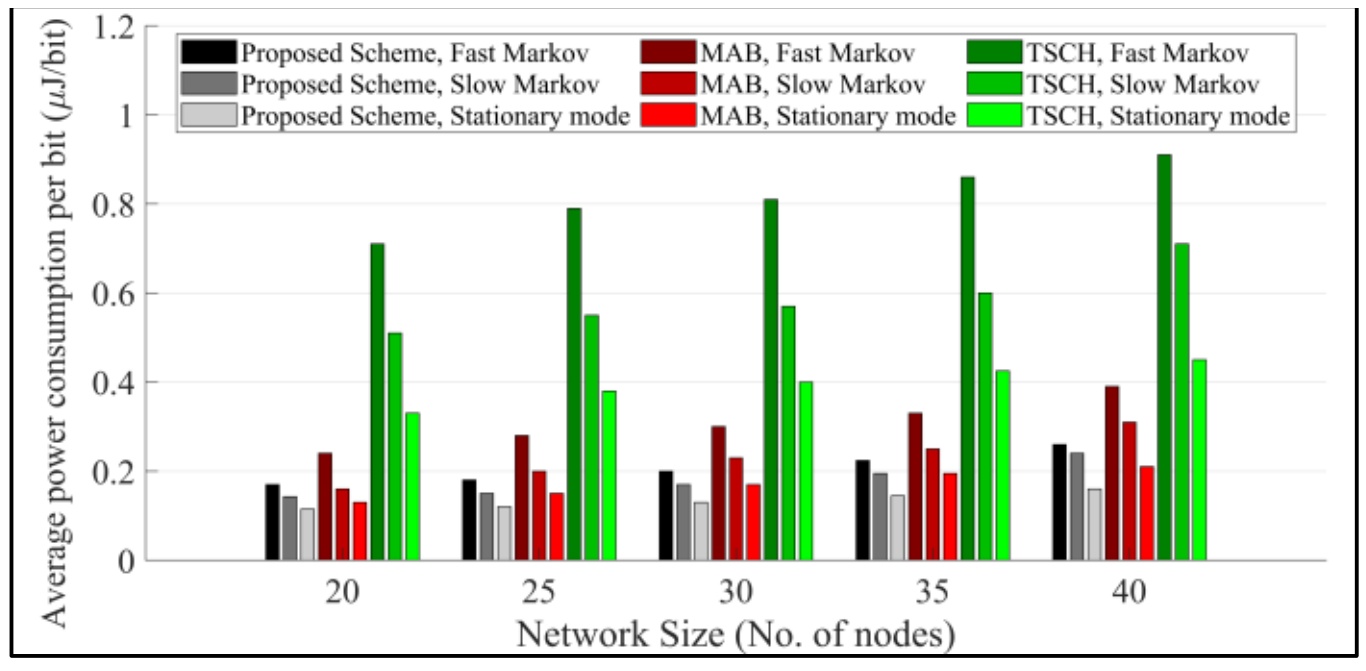

Fig. 7. Energy Efficiency vs. Network size.

### 4) *Average end-to-end delay*

In order to measure the average end-to-end delay, we focus on the worst case scenario by considering a given node and assuming that the packet is generated at the beginning of the frame. Let $\mathcal{T}_{\grave{k}\to k} = \left\{t_1^{\grave{k}\to k}, .., t_{\mathtt{m}_{\grave{k}\to k}}^{\grave{k}\to k}\right\}$ be the times-slots allocated by the scheduling subsystem to the communication link $\grave{k} \to k$. Also, we denote by $\mathtt{m}_{TX}^{\grave{k}\to k}$ the number (re)transmissions actually occurred until the packet generated by $\grave{k}$ in time-slot 0 is delivered to the next hop $k$. Given that the packet has to be queued until the next outgoing cell, the $hop_delay_{\grave{k}\to k}$ can be measured by the following formula:

$$hop_delay_{\grave{k}\rightarrow k} = \tau_{slot} \times \begin{cases} \left(\left(\left\lceil \frac{\mathfrak{m}_{TX}^{\grave{k}\rightarrow k}}{\mathfrak{m}_{\grave{k}\rightarrow k}} \right\rceil - 1\right) \times SF_{length} + t^{\grave{k}\rightarrow k}_{\mathfrak{m}_{\grave{k}\rightarrow k}} + 1\right), & \mathfrak{m}_{\grave{k}\rightarrow k} | \mathfrak{m}_{TX,\grave{k}\rightarrow k} \\ \left(\left(\left\lceil \frac{\mathfrak{m}_{TX}^{\grave{k}\rightarrow k}}{\mathfrak{m}_{\grave{k}\rightarrow k}} \right\rceil - 1\right) \times SF_{length} + t^{\grave{k}\rightarrow k}_{\mathfrak{m}_{TX}^{\grave{k}\rightarrow k} (mod\ \mathfrak{m}_{\grave{k}\rightarrow k})} + 1\right), & otherwise \end{cases}$$

Obviously, if the successful transmission occurs with a greater number of attempts, the delay increases as well. In measuring the average end-to-end delay, we use $hop_delay_{\grave{k}\rightarrow k}$ to compute the delay experienced over all the communication links along the path from the packet source to the final destination. Fig. 8 plots the delay associated with all the schemes versus the network size. A slight increasing trend is witnessed in general as the number of nodes increases. Again, in all cases, the proposed approach results in fewer number of retransmissions, thereby reducing the end-to-end latency.

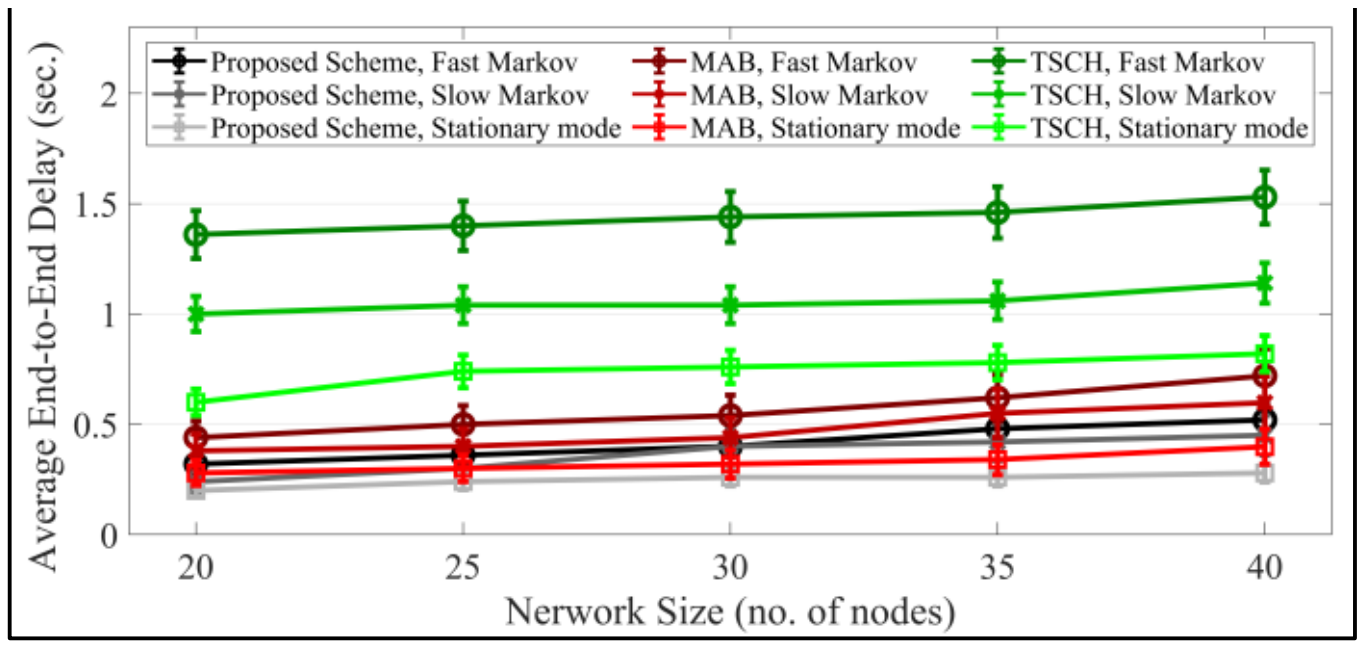

Fig. 8. End-to-end delay vs. Network size.

## VII. Experimental Setup

In this section, we briefly report on a small experiment based on a physical setup to measure the memory footprint of Algorithm 1 in an embedded IoT device. Additionally, we compare the measurements obtained from our implementation with the results of simulation under similar settings.

TABLE VIII
HARDWARE COMPONENTS USED IN THE EXPERIMENT

| Hardware Component | Manufacturer | # | Specifications |
|---|---|---|---|
| CC2650STK SensorTag | Texas Instrument | 2 | 2.4 GHz RF transceiver |
| Debugger DevPack | Texas Instrument | 2 | - |
| JN5168 dongle | NXP Semiconductors | 2 | 2.4 GHz RF transceiver<br>32-bit RISC processor, programmable clock speed<br>32 kB RAM and<br>4 kB EEPROM memory<br>256 KB memory flashed with Contiki OS v. 3.0 |

### A. Implementation setup and memory footprint

The hardware components used in our setup are listed in Table VIII. Our test consists of two miniature networks: one 2.4 GHz Bluetooth-based network which acts as external interferer and one IEEE 802.15.4 TSCH-based sender-receiver pair (See Fig. 9 for a block diagram illustration). The Bluetooth-based network consists of two CC2650STK SensorTags [49] (See Fig. 10(a)). Each tag is essentially a wireless Micro Controller Unit (MCU) which is connected to the USB port on a PC via a debugger development package (Debugger DevPack) (See Fig. 10(b)) used to program the SensorTag node [50]. The Debugger DevPack is comprised of a small XDS110 JTAG debugger with a USB connection to ensure MCU does not turn off in the midst of the testing. The 2.4 GHz transceiver on CC2650STK allows for Bluetooth or 6LowPAN communications. External interference is mimicked by these two SensorTags where one acts as master (receiver) and the other as slave (sender). The slave node generates one packet at a random point during an interval ranging from 10 msec to 80 msec to emulate high and low interference. The use of frequency band by the SensorTags is left to the default configuration in the Bluetooth frequency hopping module. As for our TSCH setup, we use a pair of USB-operated JN5168 dongles manufactured by NXP semiconductors [51] (See Fig. 11). The two dongles are flashed with Contiki OS v. 3.0 [52] which supports TSCH implementation. The DMABB-CH algorithm has been developed as a Contiki application running on the receiving dongles. After initial association, the application stands idle and only continuously listens on the UART interface. The process on the sending dongle generates one TSCH packet per timeslot. Upon reception of the very first sequence of packets, the process on the receiving dongle tunes to all physical channels according to the standard hopping sequence, and then it starts iterating over the MAIN loop in Algorithm 1 with a total of $N = 20000$ iterations.

The binary file size for Algorithm 1 developed for the receiving dongle was 10 KB, which can be read off from the SRAM usage in the .map file generated by gcc. Also, a custom routine has been developed to approximately profile the RAM usage during the runtime. At the end of the iterations, a total of 2856 bytes have been used by the program. Hence, it is possible to execute DMABB-CH on the JN5168 dongles directly. However, the maximum number of child nodes that can be supported can be within the range of 6 to 10.

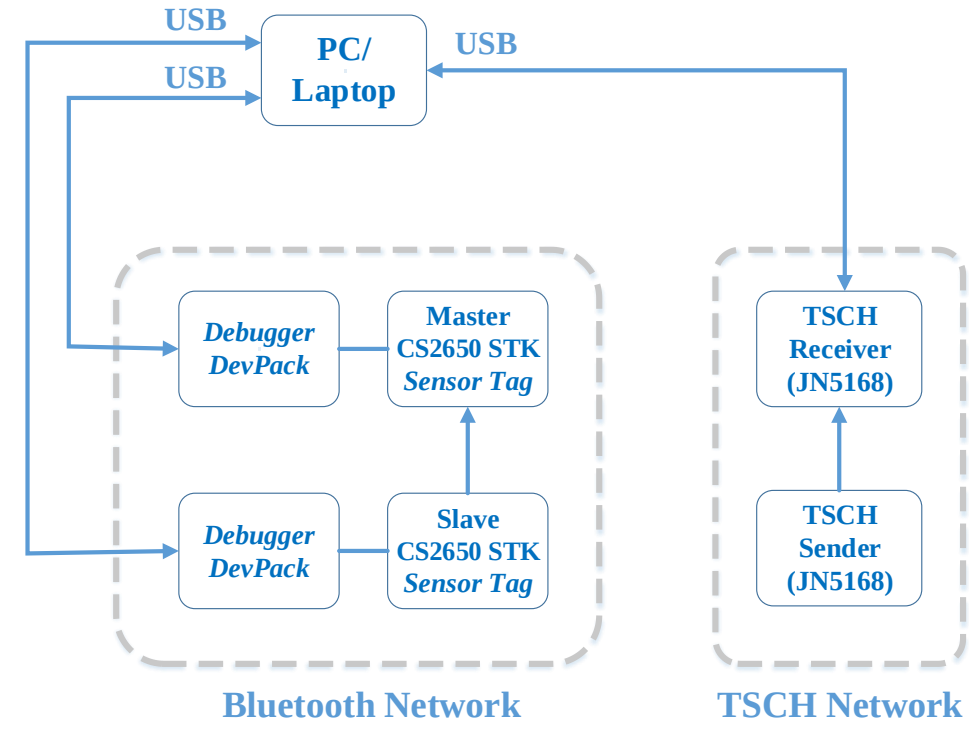

Fig. 9. Physical Setup to Measure Memory Footprint.

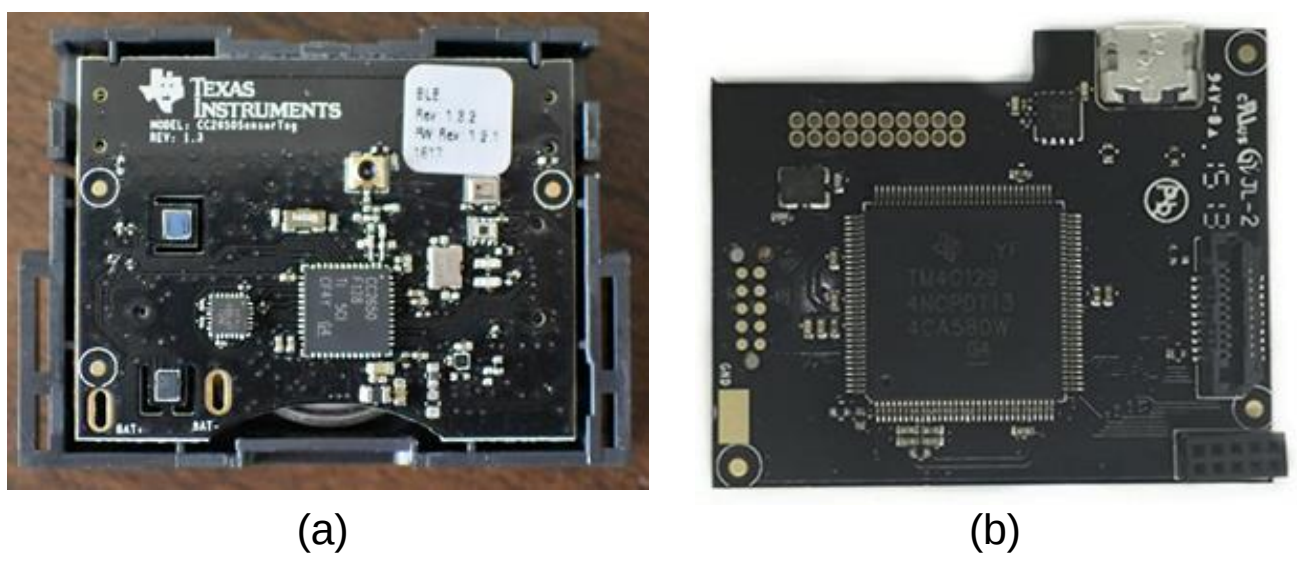

(a) (b)
Fig. 10. (a) C2650STK SensorTag. (b) Debugger DevPack.

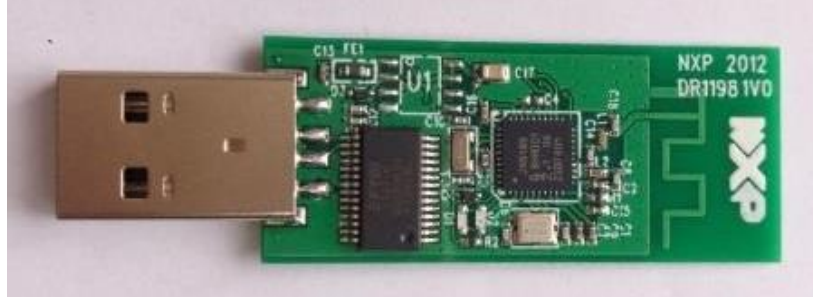

Fig. 11. NXP JN5168 USB wireless dongle.

### B. Performance measurements

In this section, we compare our hardware measurements against the simulation results. Obviously, our miniature single-hop setting cannot be used for end-to-end measurements. Alternatively, we turn our attention to more elemental metrics: link-layer *Packet Acknowledgement Ratio* (*PAR*) and *single-hop delay*. PAR is the ratio of the number of packets acknowledged by a given node's neighbor to the number of packets sent to that neighbor. This is an informative metric as it can be used as the main proxy for PDR. A low PAR indicates that several retransmissions are required to achieve high end-to-end PDR. As for single-hop delay, we argue that it is an even better metric to isolate the impact of the channel hopping strategy compared to end-to-end delay. This is because the latter is also very much affected by the scheduling algorithm.

In Section VI, we reported on NS-3 simulation results based on Wi-Fi interference. Here, to match with our physical test-bed, we need to simulate the impact of a Bluetooth interferer. The official release of NS-3 still lacks BLE support, but we used the extension in [53] as the BLE stack (BLE 4.1) for our simulation. In both the simulation and the test setup, we model the Bluetooth interference activity as a two-state ON/OFF Markov chain (See Fig. 12). In the ON state, the slave node in the Bluetooth network generates data packets with mean Poisson rate $\chi$ pkts/sec, and transmits to the master node with average TX power of 10 mW. In the plots, we report results for $\chi \in \{5,10,20\}$ pkts/sec. In the OFF state, the Bluetooth network is dormant. This simple ON/OFF model can capture non-stationary and bursty interference behavior. The 2.4 GHz channel occupancy is left to the default Bluetooth frequency hopping module. In simulation, we place the TSCH and the Bluetooth sender-receiver pairs in close proximity to mimic the test setup. The rest of the parameters for the TSCH network are chosen according to Table III for both the simulation and the physical setup. Fig. 13 and Fig. 14 show the results obtained for average PAR and average single-hop delay, respectively. The plots demonstrate the impact of the aggressiveness of the Bluetooth interference activity on the performance of our DMABB-CH algorithm as well as the two baseline schemes, and are drawn with 95% confidence level in stacked bar graph. The superiority of DMABB-CH in all cases of Bluetooth interference is consistent with the simulation results given for Wi-Fi interference in Section VI. A slight difference is noted between the values obtained from simulating Bluetooth interference and the results obtained from hardware measurements. In fact, the simulation results are more optimistic across all three schemes. This is partially caused by the blacklisting method that is used by the Bluetooth hopping module. It should be noted that blacklisting methods may differ across Bluetooth stack implementations. Moreover, although the empirical tests have been conducted in an isolated lab to ensure no outsider interference (except for the coexisting Bluetooth transmitter), generating the same context via simulation is no mean feat. It is very much dependent on the tuning of the physical layer parameters and the selected propagation models.

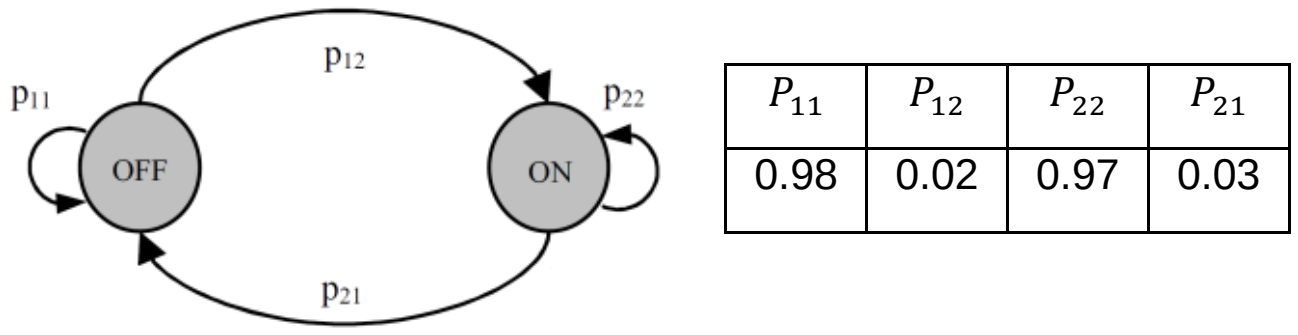

| $P_{11}$ | $P_{12}$ | $P_{22}$ | $P_{21}$ |
|---|---|---|---|
| 0.98 | 0.02 | 0.97 | 0.03 |

Fig. 12. The Bluetooth interferer ON/OFF model.

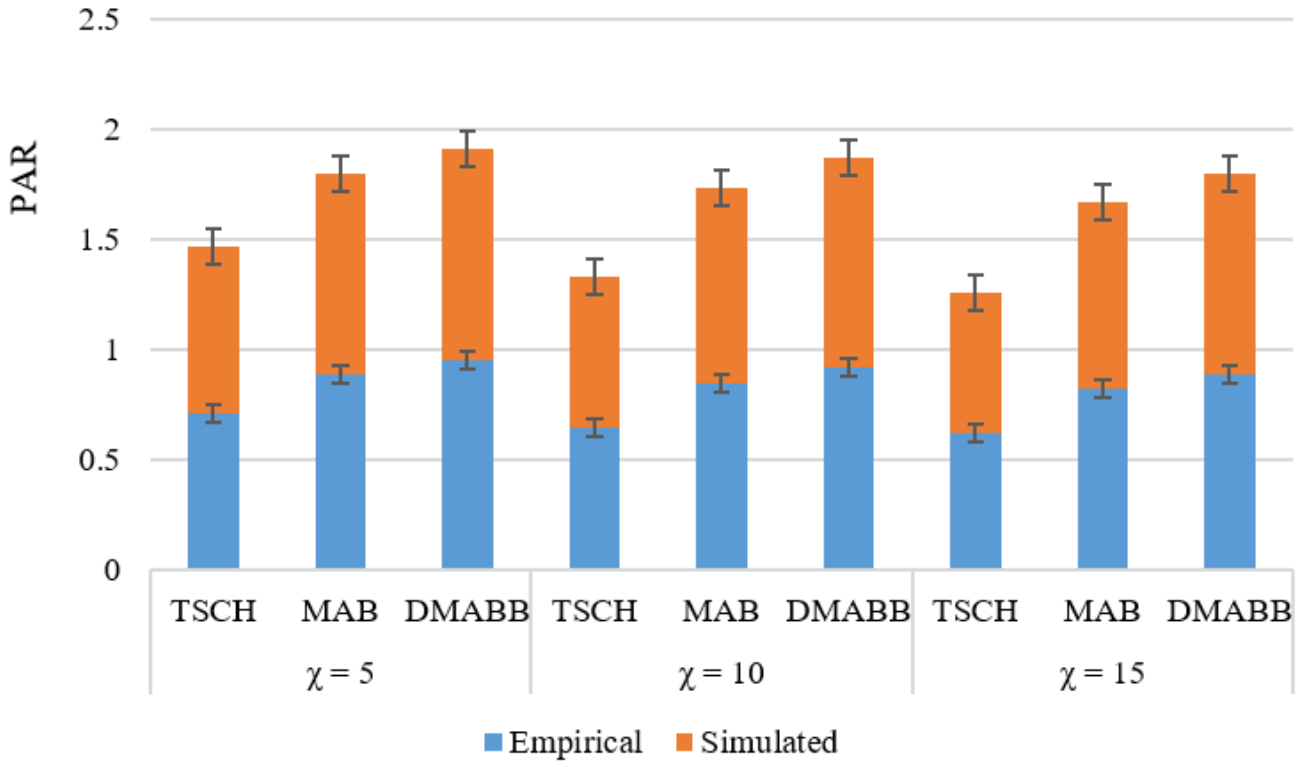

Fig. 13. Comparison of average PAR.

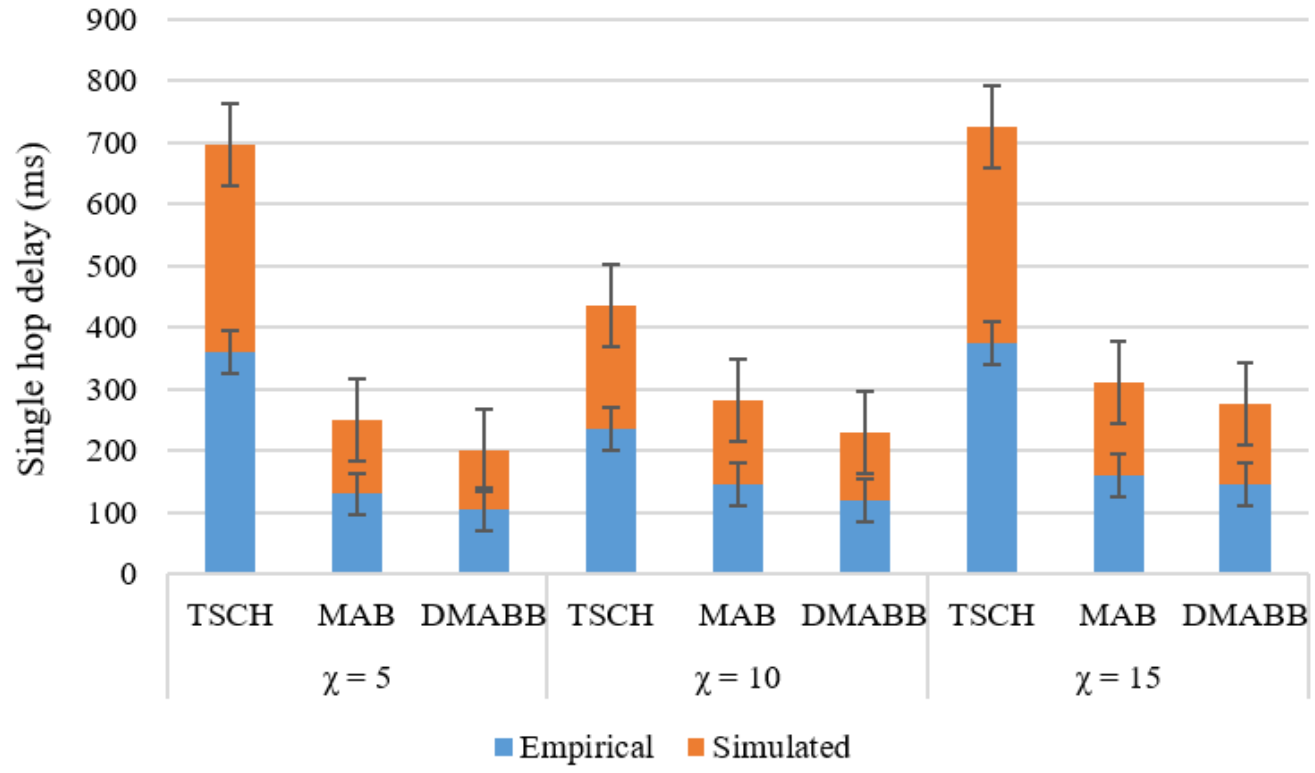

Fig. 14. Comparison of average single-hop delay.

## VIII. Conclusion

We addressed the problem of channel hopping in IEEE 802.15.4-2015 TSCH networks. Assuming that the statistics of the external interference is not known beforehand, we proposed a lightweight learning-based algorithm that can be used independently by the receiving end of each link to select near-optimal physical channels over time. The proposed algorithm has low computational complexity. Also, by implementing a physical test setup, it was shown that its memory footprint is within the confines of embedded devices used in IoT scenarios. Through simulations, the proposed scheme has been compared against default TSCH and a state-of-the-art multi-armed bandit-based scheme. As evidenced by the results, accounting for the non-stationarity of the interference can improve the network performance in terms of packet delivery ratio, thereby resulting in higher energy efficiency and lower end-to-end delay.

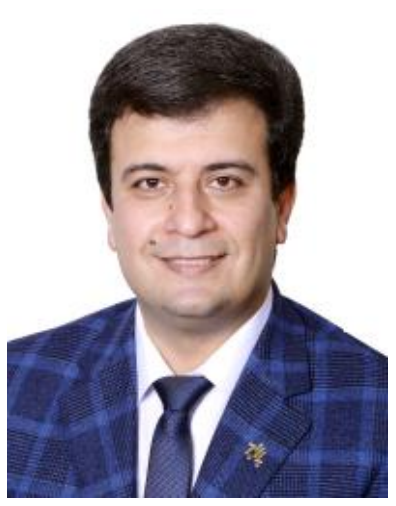

**Nastooh Taheri Javan** (S'12-M'14-SM'20) received his Ph.D. in computer engineering from Amirkabir University of Technology (Tehran Polytechnic), Tehran, Iran, in 2017. Dr. Taheri Javan is currently a post-doc fellow in the Computer Engineering department at Amirkabir University of Technology (Tehran Polytechnic). His research interests include wireless networks, network coding and optimization.

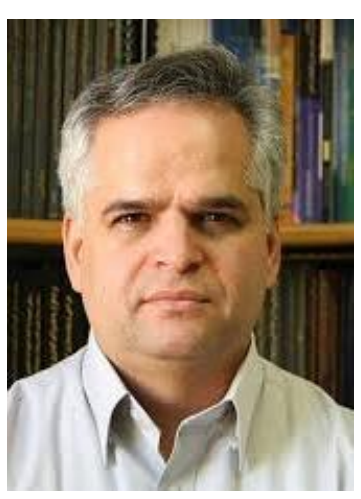

**Masoud Sabaei** is an Associate Professor with the Computer Engineering department at Amirkabir University of Technology, in Tehran, Iran. Dr. Sabaei received his Ph.D. in computer engineering from Amirkabir University of Technology, Tehran, Iran, in 2000. His research interests are wireless networks and software-defined networks.

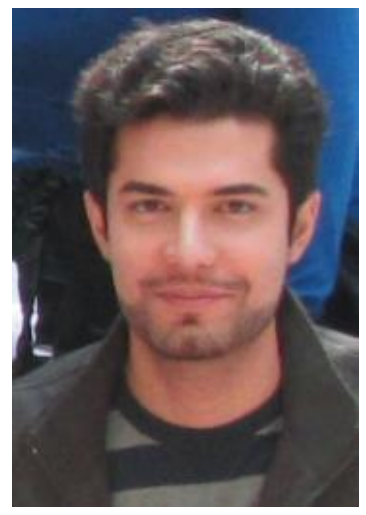

**Vesal Hakami** received his Ph.D. in computer networking from Amirkabir University of Technology, Tehran, Iran, in 2015. In 2016, Dr. Hakami joined as an Assistant Professor to the School of Computer Engineering, Iran University of Science and Technology, Tehran, Iran. His research focuses on resource control and optimization for computer networks.